\documentclass[twocolumn]{aastex631}
\usepackage{graphicx}
\usepackage{float}
\usepackage{amsmath}
\usepackage{amssymb}
\usepackage{subfigure}
\usepackage{cases}
\usepackage{mathrsfs}
\usepackage[flushleft]{threeparttable}
\usepackage{amssymb}
\usepackage{pifont}
\usepackage{epstopdf}
\usepackage{xcolor}
\usepackage[all]{hypcap}
\usepackage{mwe,tikz}
\usepackage{bm}
\usepackage{multirow}
\usepackage{longtable}
\usepackage{xspace}
\usepackage{rotating}
\usepackage{CJK}
\usepackage{url}
\usepackage{upgreek}
\usepackage{booktabs}
\usepackage{textcomp}
\usepackage{makecell}
\usepackage{enumitem}
\usepackage{csquotes}
\usepackage{soul}
\usepackage{tabularx}

\newcommand{\e}{\text{e}}

\newcommand{\p}{\partial}

\newcommand{\fermi}{{\em Fermi}\xspace}
\newcommand{\swift}{{\em Swift}\xspace}

\newcommand{\sw}[1]{\texttt{#1}}

\newcommand{\mwgrb}{GRB 241030A}
\newcommand{\mfcgrb}{GRB 211211A}

\newcommand{\bat}{BAT}
\newcommand{\xrt}{XRT}
\newcommand{\uvot}{UVOT}
\newcommand{\gbm}{GBM}

\begin{document}

\title{ Fast-Cooling Synchrotron Prompt Emission from Internal Shocks in GRB 241030A}

\correspondingauthor{Bin-Bin Zhang, Xiao-Hong Zhao}
\email{bbzhang@nju.edu.cn; zhaoxh@ynao.ac.cn}

\author[0000-0003-1471-2693]{Varun}
\affiliation{School of Astronomy and Space Science, Nanjing University, Nanjing 210093, China}
\affiliation{Key Laboratory of Modern Astronomy and Astrophysics (Nanjing University), Ministry of Education, China}
\affiliation{KTH Royal Institute of Technology, Department of Physics, 106 91 Stockholm, Sweden}

% \author[0000-0002-5596-5059]{Yi-Han Iris Yin}
% \affiliation{School of Astronomy and Space Science, Nanjing University, Nanjing 210093, China}
% \affiliation{Key Laboratory of Modern Astronomy and Astrophysics (Nanjing University), Ministry of Education, China}

\author[0000-0003-4111-5958]{Bin-Bin Zhang}
\affiliation{School of Astronomy and Space Science, Nanjing University, Nanjing 210093, China}
\affiliation{Key Laboratory of Modern Astronomy and Astrophysics (Nanjing University), Ministry of Education, China}
\affiliation{Purple Mountain Observatory, Chinese Academy of Sciences, Nanjing, 210023, China}

\author[0000-0003-3659-4800]{Xiao-Hong Zhao}
\affiliation{Yunnan Observatories, Chinese Academy of Sciences, Kunming, China}
\affiliation{Center for Astronomical Mega-Science, Chinese Academy of Sciences, Beijing, China}
\affiliation{Key Laboratory for the Structure and Evolution of Celestial Objects, Chinese Academy of Sciences, Kunming, China}

\author[0000-0002-5485-5042]{Jun Yang}
\affiliation{Institute for Astrophysics, School of Physics, Zhengzhou University, Zhengzhou 450001, China}
%\affiliation{School of Astronomy and Space Science, Nanjing University, Nanjing 210093, China}
%\affiliation{Key Laboratory of Modern Astronomy and Astrophysics (Nanjing University), Ministry of Education, China}

% \author[0000-0002-9725-2524]{Bing Zhang}
% \affiliation{Nevada Center for Astrophysics, University of Nevada Las Vegas, NV 89154, USA}
% \affiliation{Department of Physics and Astronomy, University of Nevada Las Vegas, NV 89154, USA}

\author[0009-0009-2083-1999]{Run-Chao Chen}
\affiliation{School of Astronomy and Space Science, Nanjing University, Nanjing 210093, China}
\affiliation{Key Laboratory of Modern Astronomy and Astrophysics (Nanjing University), Ministry of Education, China}

\author[0000-0002-7876-7362]{Vikas Chand}
\affiliation{Louisiana State University, Department of Physics and Astronomy, Baton Rouge, Louisiana 70803, USA}

\begin{abstract}

We present a time–resolved, joint \textit{Swift}--\textit{Fermi} spectral study of GRB\,241030A (\(z=1.411\)) that cleanly isolates the synchrotron origin of its prompt emission and favors a matter–dominated, internal–shock scenario. The light curve shows two episodes separated by a quiescent gap. Episode~I (0–45~s) is well described by a single power law with photon index \(\simeq -3/2\), consistent with the fast–cooling synchrotron slope below the peak. Episode~II (100–200~s), exhibits two robust spectral breaks: a low–energy break at \(E_{\rm b}\!\sim\!2\)–3~keV that remains nearly constant in time, and a spectral peak \(E_{\rm p}\) that tracks the flux within pulses but steps down between them. The photon indices below and above \(E_{\rm b}\) cluster around \(-2/3\) and \(-3/2\), respectively, as expected for fast-cooling synchrotron emission. The burst displays an unusually small (consistent with zero) spectral lag across GBM bands. At later times (\(\gtrsim\)230~s), the spectrum softens toward \(\sim\!-2.7\), as expected when the observing band lies above both \(\nu_m\) and \(\nu_c\). These behaviors are difficult to reconcile with a globally magnetized outflow with a decaying field, which naturally produces hard-to-soft \(E_{\rm p}\) evolution, growing \(\nu_c\), and appreciable lags. By contrast, internal shocks with a roughly steady effective magnetic field and a time-variable minimum electron Lorentz factor (equivalently, 
{e.g., a varying fraction of accelerated electrons} simultaneously account for (i) the stable \(E_{\rm b}\), (ii) the intensity-tracking yet step-down \(E_{\rm p}\), (iii) the canonical \(-2/3\) and \(-3/2\) slopes, and (iv) the near-zero lag.

\end{abstract}


%\tableofcontents

\section{Introduction}
\label{sec:intro}

 Gamma-ray burst (GRB) research has come a long way toward understanding their origins, classification, temporal properties, and spectral modeling. However, several aspects of GRB physics are still unresolved. One key topic that particularly needs clarity is the radiative processes responsible for the broadband spectrum. The most natural candidate is synchrotron radiation from energetic electrons in the presence of strong magnetic fields. The theoretical synchrotron spectrum expected from the mechanism consists of power-law segments with breaks at two characteristic frequencies ($\nu_c$ and $\nu_m$) \citep{cohenPossibleEvidenceRelativistic1997,1998ApJ...494L..49S,2000MNRAS.313L...1G}. These predictions are not fully consistent with gamma-ray observations \citep{1998ApJ...506L..23P, 2002A&A...393..409G, kanekoCompleteSpectralCatalog2006}. Typical photon indices in lower energy bands are harder ($\langle \alpha \rangle \sim -1$) than the predicted $\alpha^{\rm syn} = -1.5$ expected for the fast cooling synchrotron model, which is the so-called fast cooling problem \citep{2000MNRAS.313L...1G}. Various modifications to the standard synchrotron scenario have been proposed to alleviate this discrepancy, such as a decaying magnetic field \citep{2006ApJ...653..454P, uhmFastcoolingSynchrotronRadiation2014,2014ApJ...780...12Z}, magnetic reconnection and turbulence \citep{ZhangYan2011}, or additional radiative processes like inverse Compton scattering \citep{2001A&A...372.1071D,2009ApJ...703..675N,2011A&A...526A.110D}.

A small number of GRBs have simultaneous X-ray–to–gamma-ray coverage, providing a rare opportunity to identify both the low-energy break ($E_b$) and the spectral peak ($E_p$) in the same event, thanks to broadband observations enabled by the \swift\ observatory's \xrt\ and \bat\ instruments in conjunction with \fermi\ \gbm. These spectra often exhibit two distinct breaks in the $\nu F_{\nu}$ representation, with photon indices close to the theoretical synchrotron values of $-2/3$ and $-3/2$ \citep{2017ApJ...846..137O,2018A&A...616A.138O}. 
Such features are naturally explained if $E_b$ and $E_p$ correspond to the cooling and characteristic synchrotron frequencies ($\nu_c$ and $\nu_m$), respectively, under an approximately constant magnetic field. 
These broadband detections therefore indicate that at least a subset of GRBs can be well described by the standard fast-cooling synchrotron model, providing direct observational support for this mechanism.

The next step is to track how these break energies evolve in luminous GRBs where finer time-resolved spectroscopy is feasible. In the few cases studied so far, $E_p$ generally evolves more rapidly than $E_b$, although the detailed pattern of $E_p$ evolution differs from burst to burst. In this work, we present a time-resolved, joint \fermi--\swift\ analysis of the GRB prompt emission to track the evolution of $E_b$ and $E_p$ and to test synchrotron-based interpretations against alternatives. Section~\ref{sec:analysis} describes the observations and data reduction for \fermi\ and \swift. Section~3 characterizes the temporal properties and reports the spectral-lag measurements. Section~4 presents the time-resolved spectral modeling. Section~5 discusses the physical implications for the radiation mechanism and jet composition. We summarize our conclusions in Section~\ref{sec:sum}.

\section{Observation and Data Reduction}\label{sec:analysis}

\mwgrb\ was observed on 30 October 2024 by multiple instruments, from the optical to the gamma-ray bands. The \fermi\ Gamma-ray Burst Monitor (\gbm) was triggered on the event at 05:48:03~UT, localizing it to a region with an approximate uncertainty of $5^\circ$ \citep{2024GCN.37955....1F}. The \swift\ Burst Alert Telescope (\bat) also detected the burst at 05:48:03~UT, providing a position at $\mathrm{RA}=343.033^\circ$, $\mathrm{Dec}=+80.439^\circ$ with a $3\arcmin$ uncertainty, and promptly initiated follow-up observations by other instruments \citep{2024GCN.37956....1K}. The \xrt\ and \uvot\ began observations $\sim80$~s after the \bat\ trigger \citep{2024GCN.37962....1B,2024GCN.37974....1B} and refined the position to $\mathrm{RA}=343.13898^\circ$, $\mathrm{Dec}=+80.44974^\circ$ with a 90\% uncertainty of $2\arcsec$. Several ground-based observatories, including the Las Cumbres Observatory Global Telescope network and the Global MASTER-Net project, reported detections of a fading optical afterglow shortly after the trigger \citep{2024GCN.38220....1G, 2024GCN.37975....1L}. Spectroscopy of the optical afterglow with Keck/LRIS determined the redshift to be $z=1.411$ \citep{GCN37959}. In this work, we use prompt-emission data from the \fermi\ and \swift\ telescopes.

\subsection{\fermi-\gbm\ Data}

The GBM data for \mwgrb\ were retrieved from the Fermi Science Support Center (FSSC).\footnote{\url{https://fermi.gsfc.nasa.gov/ssc/}} For our analysis, we selected the three brightest NaI detectors with the smallest viewing angles relative to the burst direction—n0, n1, and n6—along with one BGO detector, b1. The viewing angles for these detectors were $4^\circ$ (n0), $27^\circ$ (n1), $39^\circ$ (n6), and $100^\circ$ (b1) from the GRB direction. Detectors with smaller viewing angles are better aligned with the source, yielding higher signal-to-noise ratios and more precise photon measurements \citep{Meegan2009}.

To capture the burst’s temporal structure and energy dependence, we constructed energy-resolved light curves across multiple energy bands with a time resolution of 1~s. We also applied the Bayesian Blocks algorithm \citep{Scargle2013} to adaptively bin the light curves. Unlike fixed-width binning, Bayesian Blocks identifies statistically significant changes in the signal—such as sudden rises or drops—by maximizing a fitness function over possible segmentations, providing a robust, data-driven representation of the variability.

\subsection{\swift\ BAT and XRT Data}

The \swift\ data for \mwgrb\ were retrieved from the UK \swift\ Science Data Center\footnote{Swift UK Archive: \url{https://www.swift.ac.uk/index.php}}. \bat\ light curves with 1~s time bins were generated using \sw{HEASOFT-6.34}, \sw{FTOOLS}, and the procedures described in the \swift--BAT software guide.\footnote{\swift--BAT guide: \url{https://swift.gsfc.nasa.gov/analysis/bat_swguide_v6_3.pdf}} We applied a gain correction with \sw{bateconvert}; then \sw{batbinevt} was used to produce light curves after creating a detector plane image (DPI), identifying problematic detectors, removing hot pixels, and performing mask-weighting and background subtraction with \sw{batdetmask}, \sw{bathotpix}, \sw{batmaskwtevt}, and \sw{batbinevt}, respectively. Background subtraction with coded-aperture detectors improves the signal-to-noise ratio and enables precise light-curve extraction.

The same steps used to generate light curves were followed to extract the BAT spectrum, with additional use of \sw{batphasyserr} and \sw{batupdatephakw} to compensate residual response features and ensure accurate burst positioning in instrument coordinates. The detector response matrix (DRM) was generated with \sw{batdrmgen}. These \swift\ products enable a detailed joint temporal and spectral analysis, together with the \fermi\ data, of the prompt emission from \mwgrb.

\xrt\ light-curve and spectral products were obtained from the online \swift--\xrt\ page for this GRB.

\section{Temporal Analysis}

\subsection{Light curve properties}

\begin{figure*}
 \centering
 \includegraphics[width=0.8\linewidth]{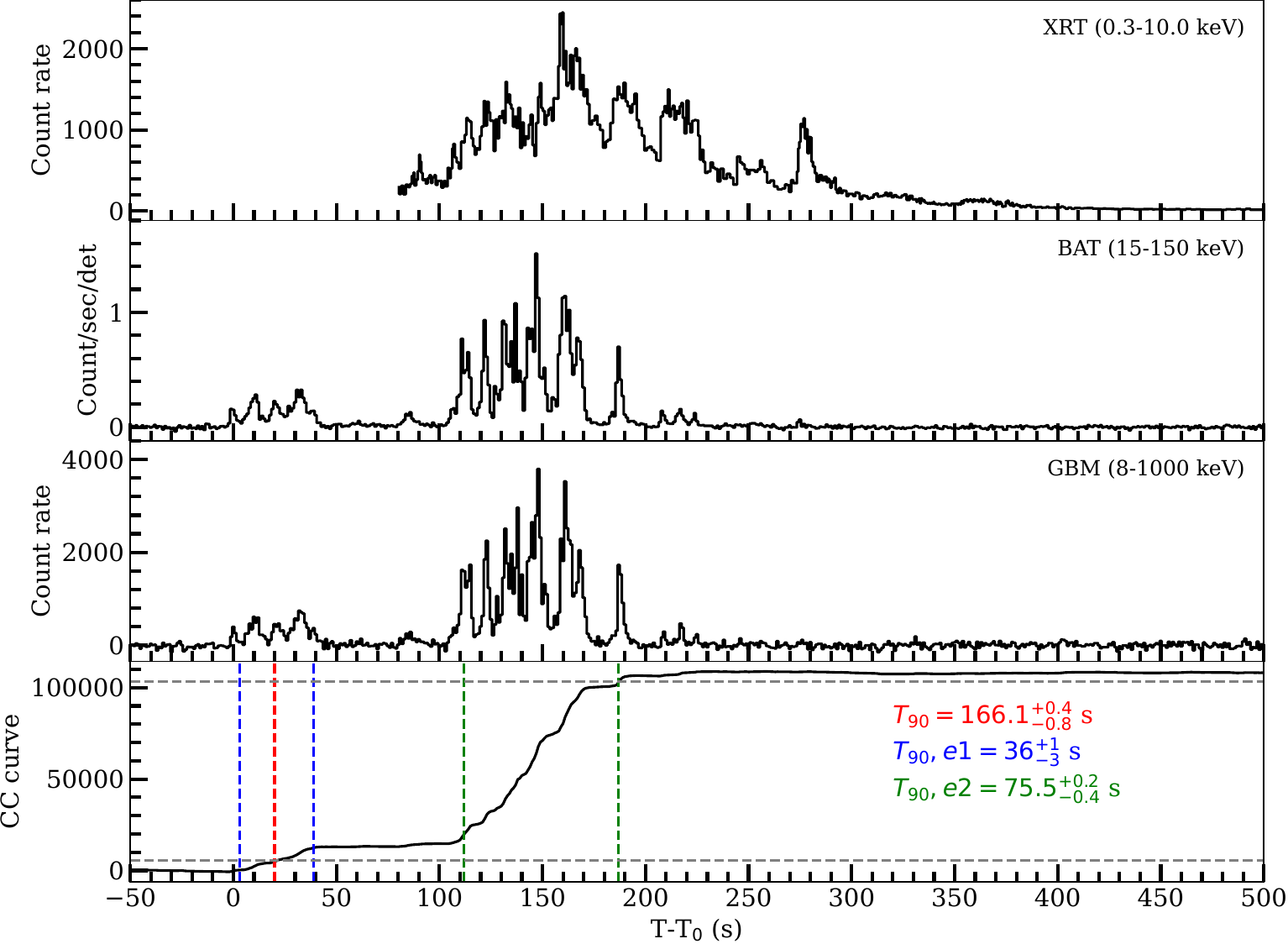}
 \caption{\mwgrb\ light curves during prompt emission as observed by \swift\ and \fermi\ telescopes. The top two panels show the soft X-ray (0.3-10 keV) and hard X-ray (15-350 keV) light curves of \xrt\ and \bat\ instruments onboard \swift\ satellite. \xrt\ light curve starts from 80 s when it slewed to the position of the event. Next panel show the \fermi/\gbm\ light curve in the energy band 8-1000 keV while its cumulative counts (CC) curve is shown underneath it in the bottom panel. A bin size of 1 sec is used for all three light curves. $T_{90}$ intervals for the full burst, its first and second episodes are marked in red, blue and green colors respectively. }
 \label{fig:lcs}
\end{figure*}

\begin{figure*}
 \centering
 \includegraphics[width=0.45\linewidth]{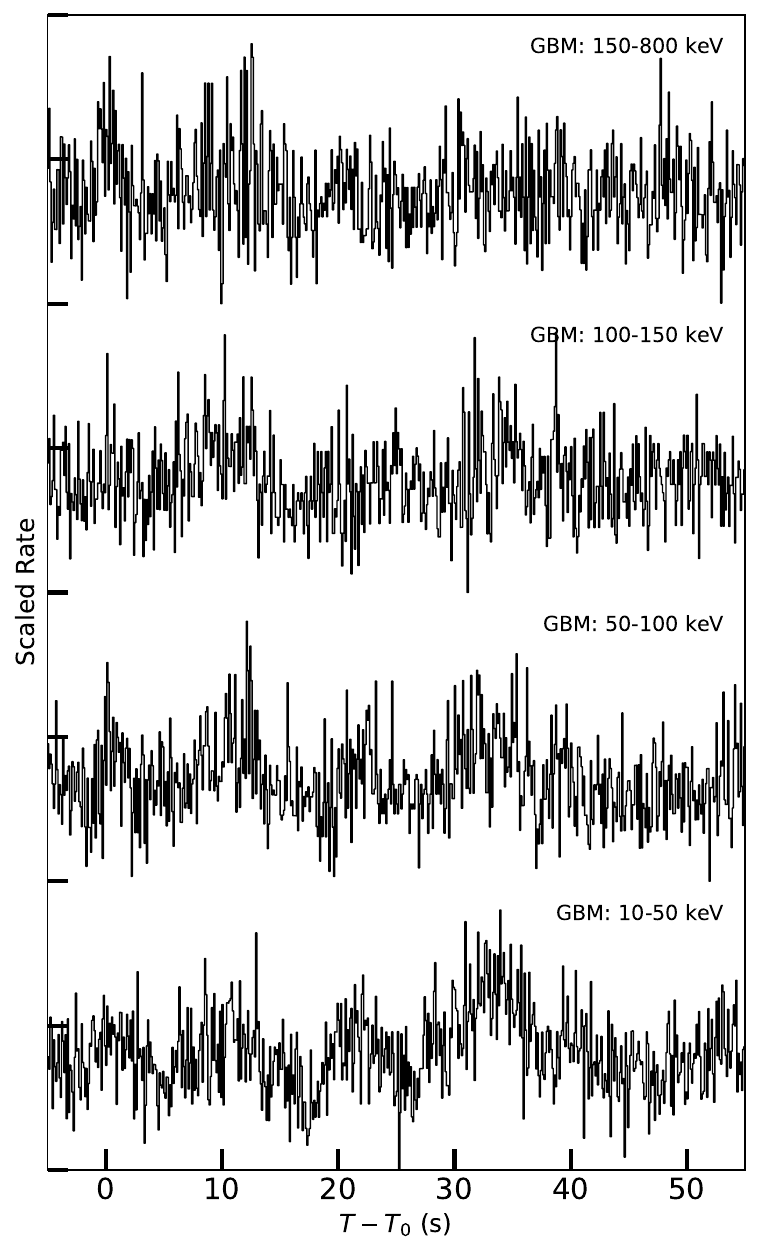}
 \includegraphics[width=0.45\linewidth]{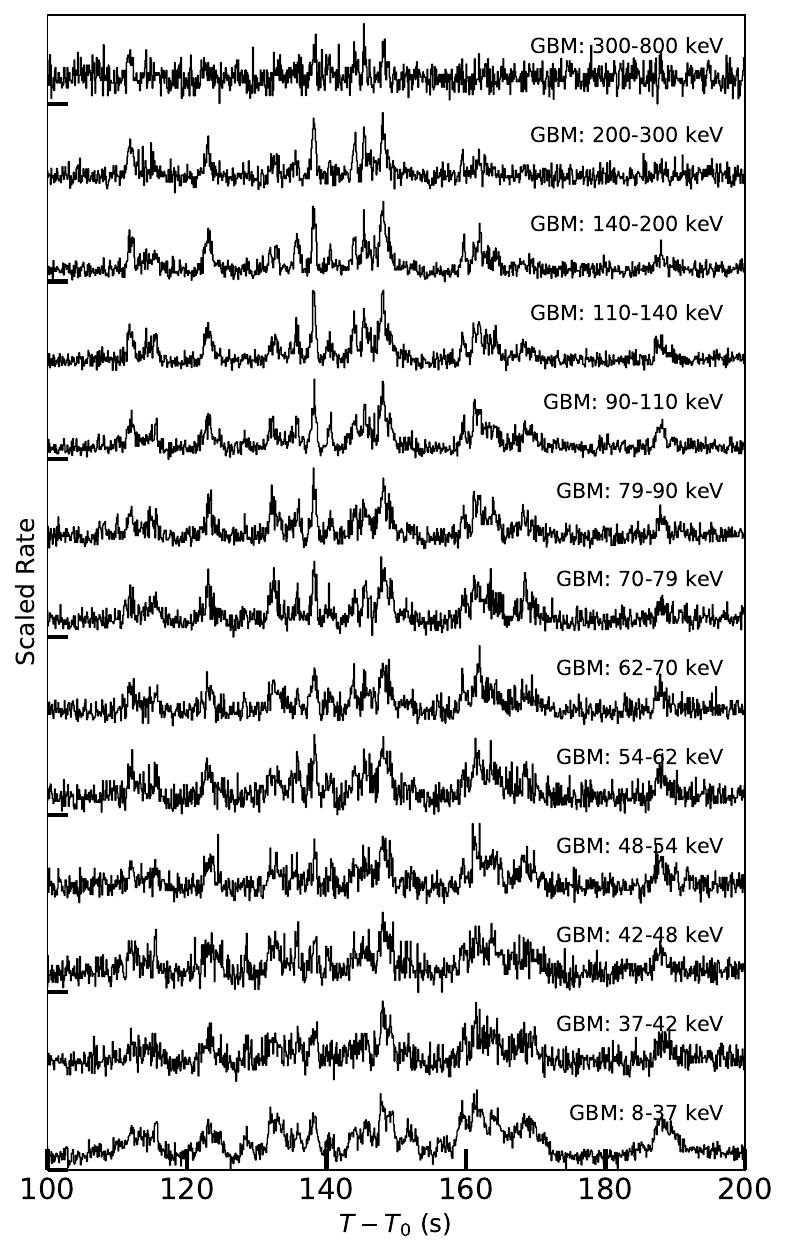}
 \caption{Energy-resolved light curves from combined selected GBM detectors from the first (left) and second (right) episodes of \mwgrb. Due to faintness of burst during the first episode, GBM data was divided in 5 energy bands whereas good statistics were obtained in 13 energy band during the second episode. Light curves have been scaled for visual clarity.}
 \label{fig:lclg}
\end{figure*}

\begin{figure*}
 \centering
 \includegraphics[width=0.45\linewidth]{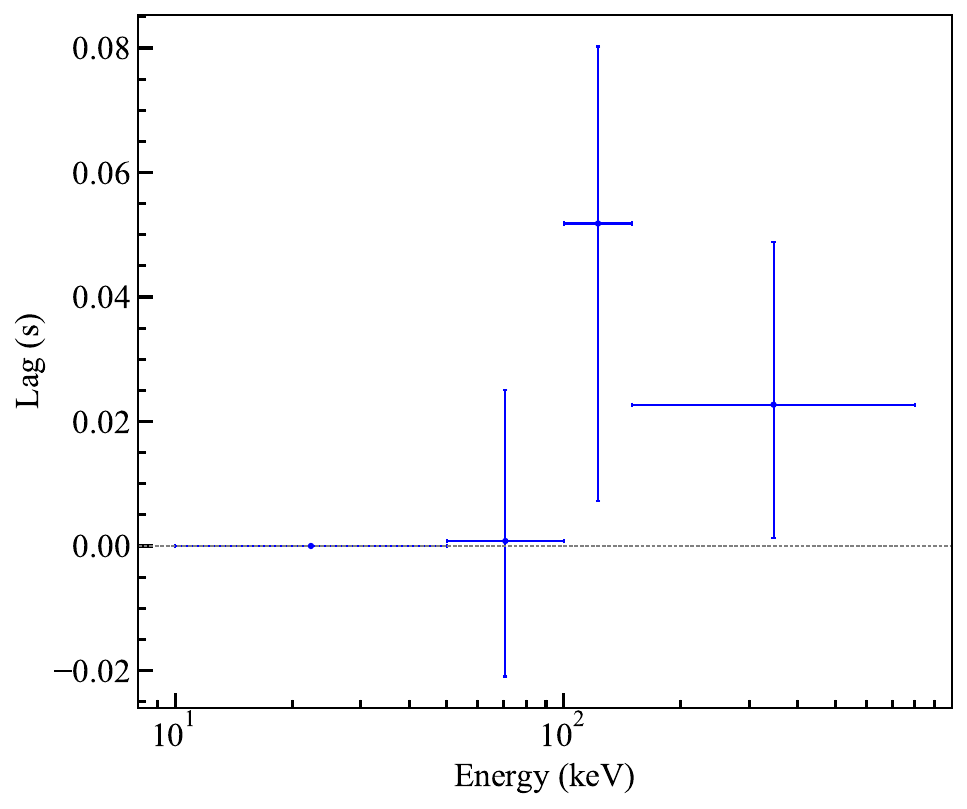}
 \includegraphics[width=0.45\linewidth]{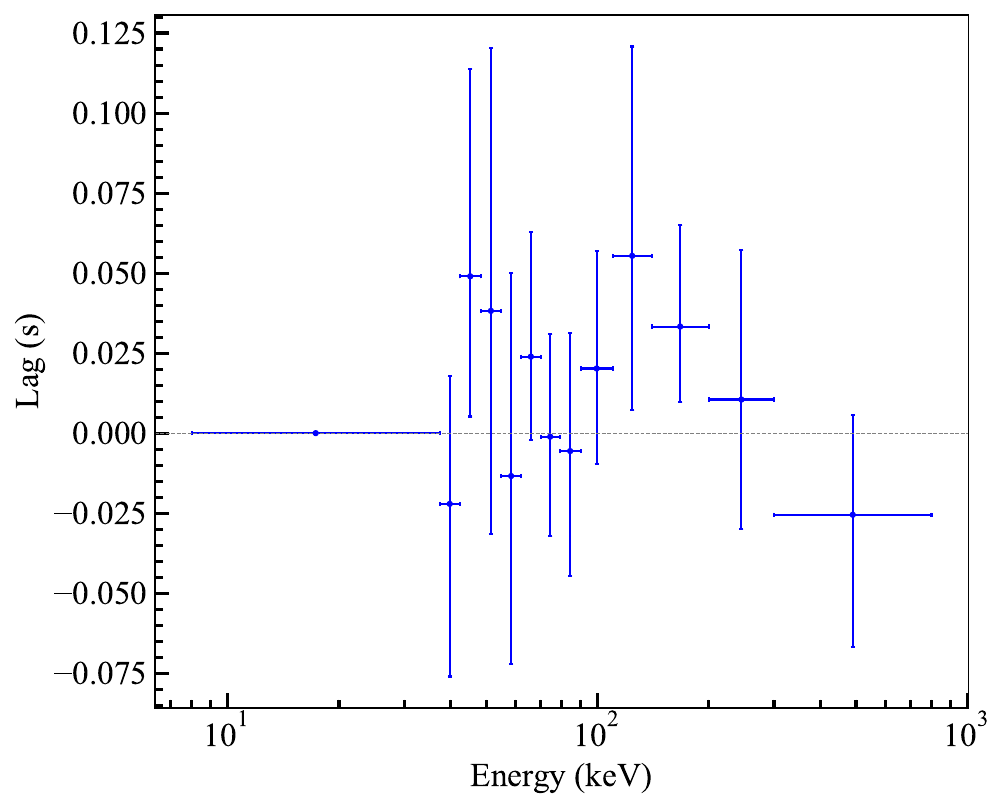} 
 \caption{Energy dependent spectral lags between light curves from the first and second episodes of \mwgrb. Positive values means soft energy photons arrive later than hard energy photons and vice versa. Error bars represent uncertainities at 1 $\sigma$ level.}
 \label{fig:spl}
\end{figure*}

\begin{figure*}
 \centering
 \includegraphics[width=0.5\linewidth]{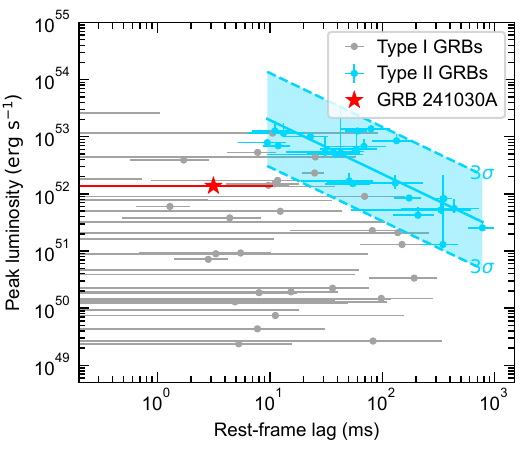}
 \caption{Placement of \mwgrb\ in the peak luminosity versus rest frame spectral lags of GRBs. Type-I and type-II GRB population data points are shown in grey and cyan dots respectively. Best-fit correlations and 3$\sigma$ variation in the data for type II GRBs are shown by solid cyan line and shaded areas respectively. }
 \label{fig:tau-L}
\end{figure*}

The X-ray and gamma-ray light curves of \mwgrb\ are shown in Figure~\ref{fig:lcs} with 1~s bins. The total prompt emission, which lasts for $\sim$230~s, is fully covered by \bat\ and \gbm. The burst comprises two distinct episodes separated by a $\sim$30~s quiescent interval. The first episode is weak, spans 0--45~s, and shows four distinct peaks. The second episode begins at $\sim$70~s and lasts until $\sim$230~s, with strong emission from 100 to 200~s consisting of six peaks. The soft X-ray emission in 0.3--10~keV observed by \xrt\ covers only the second episode, showing a slightly different, broadened peak structure and emission extending to $\sim$300~s. The duration of the full burst in terms of $T_{90,\gamma}$ (10--1000~keV) is $166.1^{+0.4}_{-0.8}$~s; for the two individual episodes it is $36^{+1}_{-3}$~s and $75.5^{+0.2}_{-0.4}$~s, respectively.

We constructed Bayesian Blocks using the method of \citet{Scargle2013} from the \gbm\ time-tagged event data in 10--1000~keV from the n0 detector (the closest to the source direction). From these blocks, we computed the minimum variability timescale (MVT), defined as half the length of the shortest block. We find an MVT of 1.1~s for the first episode and 0.14~s for the second episode. The relatively long MVT of 1.1~s in the first episode may be attributed to its faintness, which limits the resolution of faster variability; thus, a larger MVT in the first episode does not necessarily imply a larger emission radius.

\subsection{Spectral lag}

A prominent feature of GRB prompt emission is the spectral lag—the systematic delay of low-energy photons relative to high-energy photons across multi-band light curves \citep{2000ApJ...534..248N, 2006Natur.444.1044G, 2009ApJ...703.1696Z}. We searched for spectral lags in this GRB by constructing energy-resolved light curves from the \gbm\ data. Lags were measured using the cross-correlation function (CCF) method \citep{2000ApJ...534..248N,Ukwatta_2010}, and uncertainties were estimated via Monte Carlo simulations (see \citealt{Zhang_2012} for methodological details).

We first examined the two emission episodes separately (energy-resolved light curves shown in Figure~\ref{fig:lclg}) using light curve with 0.02~s bins. For episode 1, we divided the data in five bands to obtain good statistical quality: (1) 10–50~keV, (2) 50–100~keV, (3) 100–150~keV, (4) 150–200~keV, and (5) 200–900~keV. For the second episode we extracted energy-resolved light curve in 13 bands: (1) 8-37~keV, (2) 37-42~keV, (3) 42-48~keV, (4) 48-54~keV, (5) 54-62~keV, (5) 62-70~keV, (6) 70-79~keV, (7) 79-90~keV, (8) 79-90~keV, (9) 90-110~keV, (10) 110-140~keV, (11) 140-200~keV, (12) 200-300~keV, (13) 300-800~keV. We found that the resultant lags are consistent with zero in both episodes (Figure~\ref{fig:spl}). We then computed lags for individual pulses in the second episode (100–118~s, 118–128~s, 128–142~s, 142–155~s, 155–180~s, and 180–195~s). No significant lags were detected for any individual pulse.

We also examined the position of this burst in the $\tau$–$L_{\rm p}$ plane, where $\tau$ is the spectral lag and $L_{\rm p}$ is the isotropic peak luminosity, 
which are known to follow an inverse correlation in long GRBs \citep[e.g.,][]{2000ApJ...534..248N} plane in comparison with other GRBs. As shown in Figure~\ref{fig:tau-L}, \mwgrb\ lies in the region typically populated by short GRBs, with spectral lags consistent with zero and well below the $\tau$–$L_{\rm iso}$ correlation established for long GRBs \citep[e.g.,][]{2000ApJ...534..248N}. 
This shows that \mwgrb\ exhibits exceptionally small spectral lags compared with the majority of long GRBs.

\section{Spectral Analysis}

We carried out spectral fitting of this GRB using the Python package \sw{BAYSPEC}\footnote{\url{https://github.com/jyangch/bayspec}}, a Bayesian–inference–based tool for high-energy astrophysical data. For the first episode (0–45~s), we performed joint fits to the \bat\ and \gbm\ data; for the second episode (100–200~s), \xrt\ was also included. Model comparison was quantified with the Bayesian information criterion (BIC) \citep{bic_ref}. For \xrt, \bat, and \gbm\ we adopted the appropriate likelihood statistics: CSTAT \citep{Cash1979ApJ}, GSTAT \citep{Feigelson_Babu_2012}, and PGSTAT \citep{1996ASPC..101...17A}, respectively, as required by the source and background treatments of each instrument.

To characterize the emission mechanisms, we fitted the extracted spectra with several models, including standard forms such as the cutoff power law (CPL), blackbody plus cutoff power law (BB+CPL), and smoothly broken power law (SBPL). In addition, we employed a smoothly broken power law with a high-energy cutoff (CSBPL), a non-standard model well suited to the curved spectra often seen in GRBs. The CSBPL introduces a smooth transition between two power laws and an exponential cutoff at high energies, effectively capturing the observed curvature. Its functional form is
\begin{equation}
\begin{split}
N(E) &= A\,E_{\text{b}}^{\alpha_1}
\left[
\left(\frac{E}{E_{\text{b}}}\right)^{-{\alpha_1}n}
+ \left(\frac{E}{E_{\text{b}}}\right)^{-{\alpha_2}n}
\right]^{-1/n}
\\ &\quad \times \exp\!\left(-\frac{E}{E_{\text{c}}}\right),
\end{split}
\end{equation}
with
\begin{equation}
E_{\text{c}} = \frac{E_{p}}{2+\alpha_2}.
\end{equation}
Here, $A$ is the normalization; $E_{\text{b}}$ is the low-energy break where the slope changes; $\alpha_1$ and $\alpha_2$ are the photon indices below and above $E_{\text{b}}$; $n$ controls the smoothness of the break (fixed to 5.38 in our fits); $E_p$ is the high-energy break corresponding to the spectral peak in the $E F_E$ representation; and $E_{\text{c}}$ is the cutoff energy where exponential suppression begins.

First, we performed spectral analysis in coarse time bins selected to follow the distinctive peaks in the two episodes of this GRB. For the four peaks of Episode~I, the CPL model described the 10–40{,}000~keV spectra well, with statistic/dof $\sim 0.9$–$1.0$. The spectral index $\alpha_2$ lay between $-1.18$ and $-1.58$, while the peak energy $E_p$ ranged from 39 to 186~keV. 

For the early part of Episode~II (100–230~s), the 0.3–40{,}000~keV spectra were fitted with three single–component models: CPL, SBPL, and CSBPL. We included two absorption components: a Galactic component fixed at $1.79\times10^{-22}\ \mathrm{cm}^{-2}$ and an intrinsic component at the source redshift $z\sim1.4$ \citep{GCN37959}. Among these, CSBPL provided the best description. For example, in the 100–118~s interval the statistic/dof improved to $\sim 1.3$ for CSBPL, down from $\sim 1.5$ for the other two models, with similar improvements in the remaining time bins. We further found that the intrinsic absorption was consistently very low ($\sim 10^{-20}\ \mathrm{cm}^{-2}$) when using CSBPL; we therefore excluded this component from the model.

\begin{figure*}
 \centering
 \includegraphics[width=0.85\linewidth]{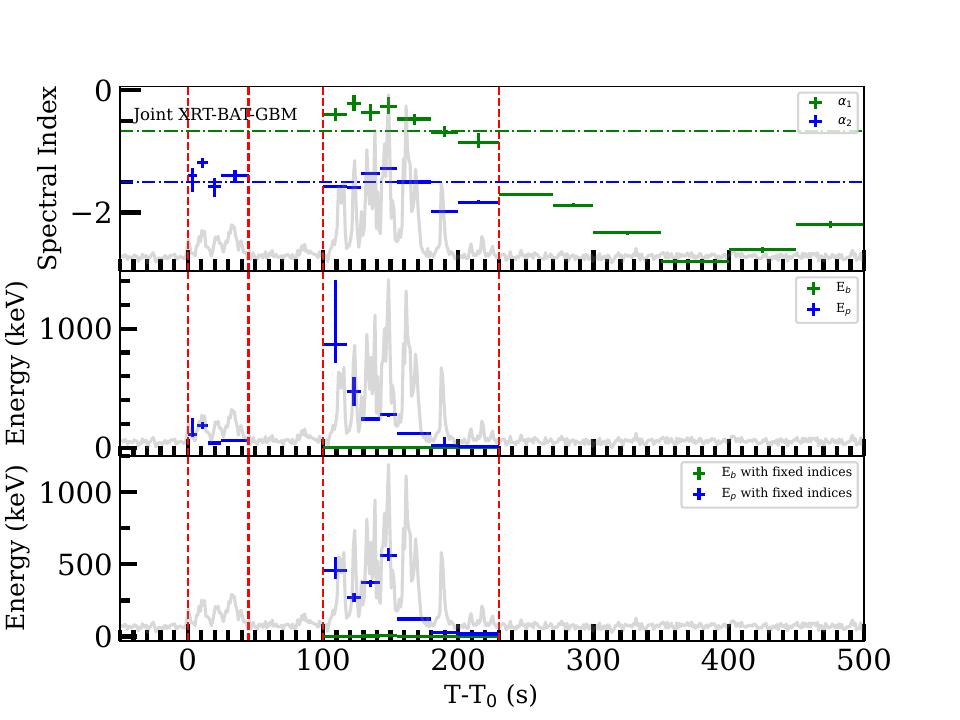}
 \caption{ Evolution of different parameters from joint fitting in coarse spectral bins. Top panel shows the evolution of photon indices $\alpha_1$ and $\alpha_2$ while the middle panel illustrates the behavior of two break energies $E_b$ and $E_p$. Dash-dotted green and blue horizontal lines in the top panel indicate the theoretically expected values of photon indices $\alpha_1$=-0.67 and $\alpha_2$=-1.5. Bottom panel show the the value of breaks at different times with indices fixed at expected values. Light curve from the nearest \gbm\ detecter to source direction is shown as grey background in all panels. }
 \label{fig:paevo}
\end{figure*}

With CSBPL, the low-energy spectral index $\alpha_1$ was between $-0.69$ and $-0.22$, the low-energy break $E_b$ remained near $\sim 2.5$~keV, and the high-energy index $\alpha_2$ ranged from $-1.98$ to $-1.28$. We also observed a systematic decrease in $E_p$ from $\sim 870$~keV at the first peak down to $\sim 21$~keV by the sixth peak of Episode~II. During the late portion of Episode~II (230–500~s), the emission was too weak at high energies, leaving only \xrt\ measurements with adequate signal. The spectra in this interval are well fitted by a simple power law, remaining relatively hard (photon index $\sim -1.8$) until $\sim 300$~s and softening to $\sim -2.7$ thereafter. The evolution of key spectral parameters is shown in Figure~\ref{fig:paevo}, and the full set of values with uncertainties is listed in Table~\ref{tab:cbin}.

\begin{figure*}
 \centering
 \includegraphics[width=0.5\linewidth]{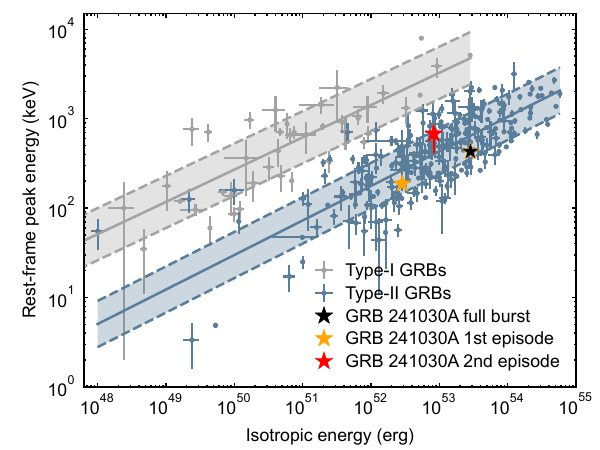}
 \caption{Placement of \mwgrb\ in the E$_{peak}$ (rest-frame peak energy) Vs. E$_{iso}$ (isotropic energy) correlation plot for GRBs. Data from two types of GRB populations are shown in grey (type-I) and cadetblue (type-II) colors respectively. Solid lines and shaded areas represent best-fit correlations and 1$\sigma$ variation in the data respectively. }
 \label{fig:amati}
\end{figure*}

In addition to the pulse-integrated fits, we also fit the episode-integrated spectra to capture the phase-averaged properties (see Table~\ref{tab:cbin}). For Episode~I, the spectrum is well described by the CPL model with a photon index of $\alpha_2 \sim -1.4$ and a peak energy $E_{\rm p} \sim 80$~keV. 
Episode~II (100–230~s) is better fitted by the CSBPL model, yielding $\alpha_2 \approx -0.2$, $\alpha_2 \approx -1.6$, and a peak energy of $E_{\rm p} \sim 280$~keV and a cut-off energy of $E_{\rm b} \sim 2$ ~keV. 
We further examined the locations of both episodes in the $E_{\rm p,i}$–$E_{\rm iso}$ (Amati) plane to test their consistency with the global correlation observed in long GRBs (Figure~\ref{fig:amati}). 
Both episodes fall within the 1$\sigma$ scatter of the Amati relation, suggesting that they follow the same spectral–energetic trend typically found in long GRBs.

\begin{figure*}
 \centering
 \includegraphics[width=0.85\linewidth]{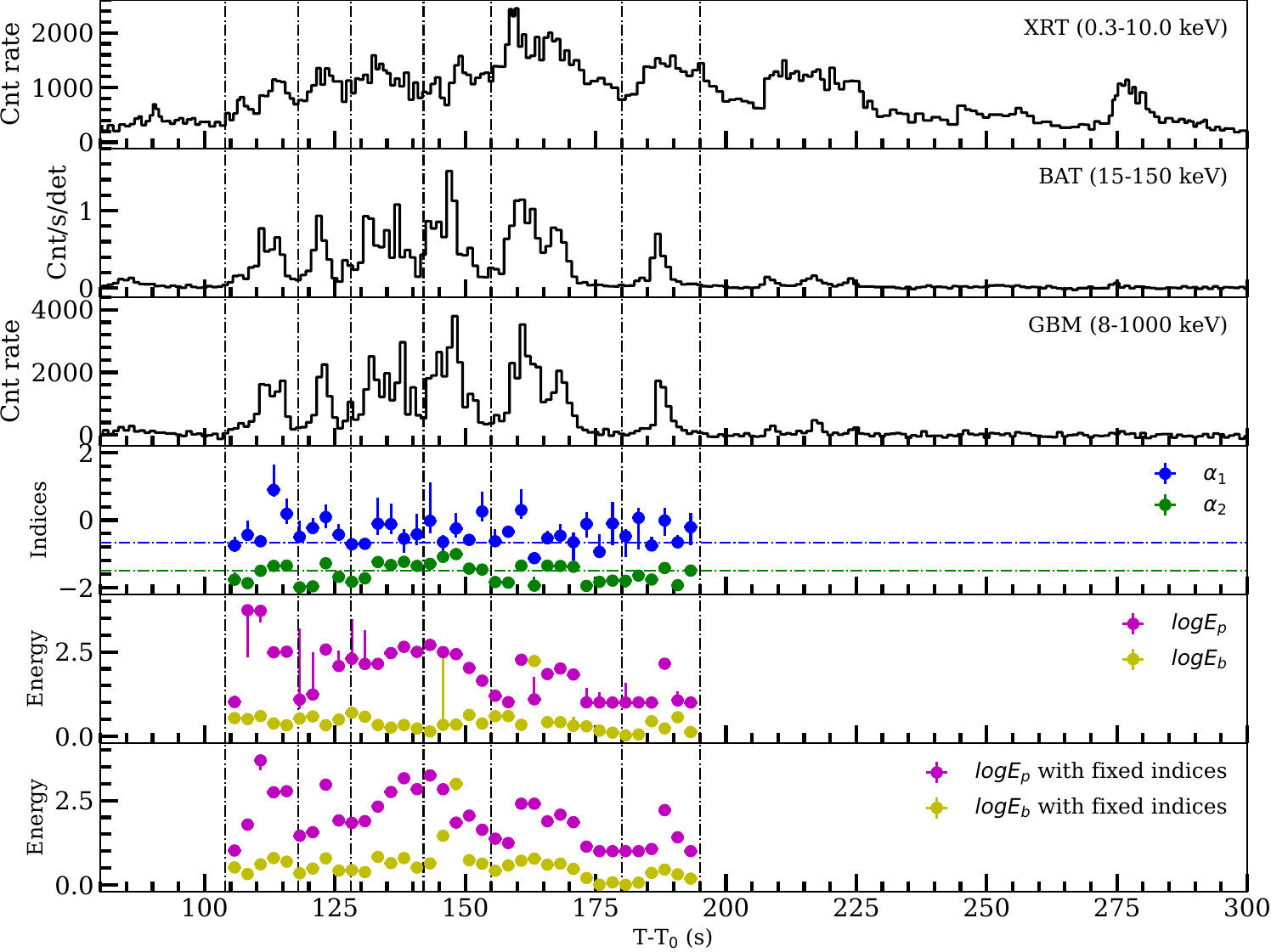}
 \caption{Evolution of spectral parameters using finer time bins spectral data. Top three panels show light curves from \xrt, \bat, and \gbm\ instruments. Next two panels show the behavior of photon indices ($\alpha_1$ and $\alpha_2$) and spectral breaks ($E_b$ and $E_p$). Dash-dotted blue and green horizontal line in photon indices panel (4th from top) indicate the mean values of photon indices. In the bottom panel, values of spectral breaks are obtained by fixing photon indices at their respective mean values.}
 \label{fig:fiev}
\end{figure*}

Next, we divided the spectral data into finer bins. For Episode~II, we selected intervals where the burst was reasonably bright, excluding data before 104~s and after 195~s. We also omitted 172–179.5~s due to poor statistics. Within the remaining range, we used 2.5~s time slices and fitted each spectrum with the CSBPL model, identified as the best continuum model from the coarse binning. The resulting evolution of key parameters is shown in Figure~\ref{fig:fiev}, with detailed values listed in Table~\ref{tab:spec_fit1}.

The low-energy index $\alpha_1$ varies between 0 and $-1$ (aside from a few outliers), with a mean $\langle \alpha_1 \rangle = -0.28$. The index $\alpha_2$, which describes the slope below the peak energy, is softer than the typical value $\simeq -1$, i.e., we find $-2 < \alpha_2 < -1$ with a mean $\langle \alpha_2 \rangle = -1.55$. The low-energy break $E_b$ remains nearly constant at $\sim 2.5$~keV, while the high-energy break $E_p$ shows an overall decrease. Within individual pulses, $E_p$ generally tracks the flux, with the spectral peak broadly following the pulse intensity. These trends become even clearer when the spectral indices are fixed at their mean values, as shown in the bottom panel of Figure~\ref{fig:fiev}.

Besides the single-component models, we also examined whether the inclusion of an additional thermal component could improve the spectral fits.
The broad 0.3–1000~keV spectra of \mwgrb\ have been modeled with a blackbody (BB) thermal component plus a cutoff power law (CPL) nonthermal component \citep{2025arXiv250104906W}. We compared this two–component model with the single–component models used in this work. Resulting parameter values from the spectral fitting are listed in Table~\ref{tab:spec_fit2}. 
In time–resolved fits, both approaches provide good descriptions of the data with comparable BIC values. Some time slices are better fit by a single component, whereas others are slightly favored by the two–component model. We therefore conclude that, statistically, both model families fit the spectra equally well. Given this and in the absence of independent evidence for an additional thermal component, we find no compelling reason to introduce a second component for \mwgrb\ in our analysis.

\begin{figure*}
 \centering
 \includegraphics[width=0.45\linewidth]{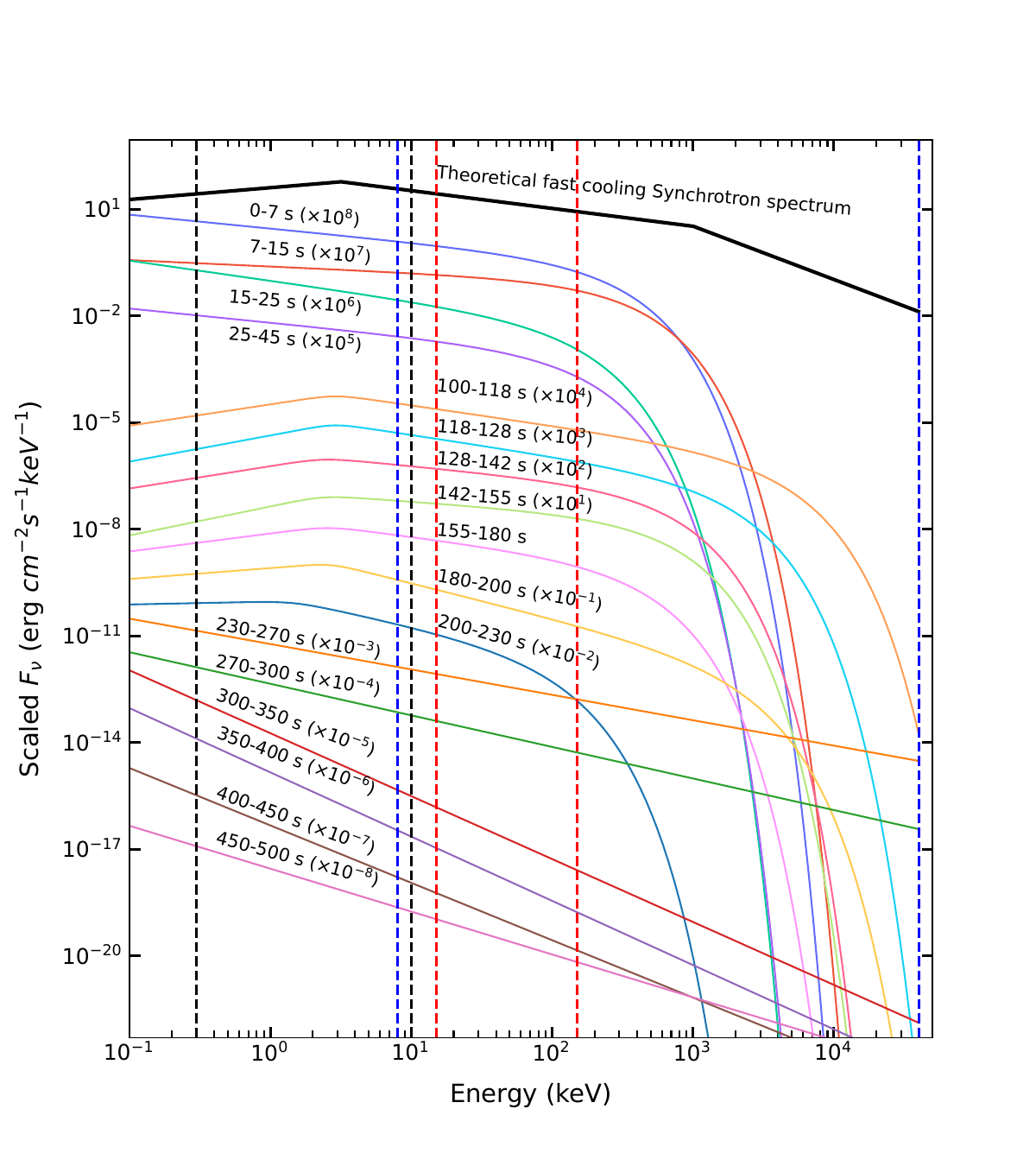}
 \includegraphics[width=0.45\linewidth]{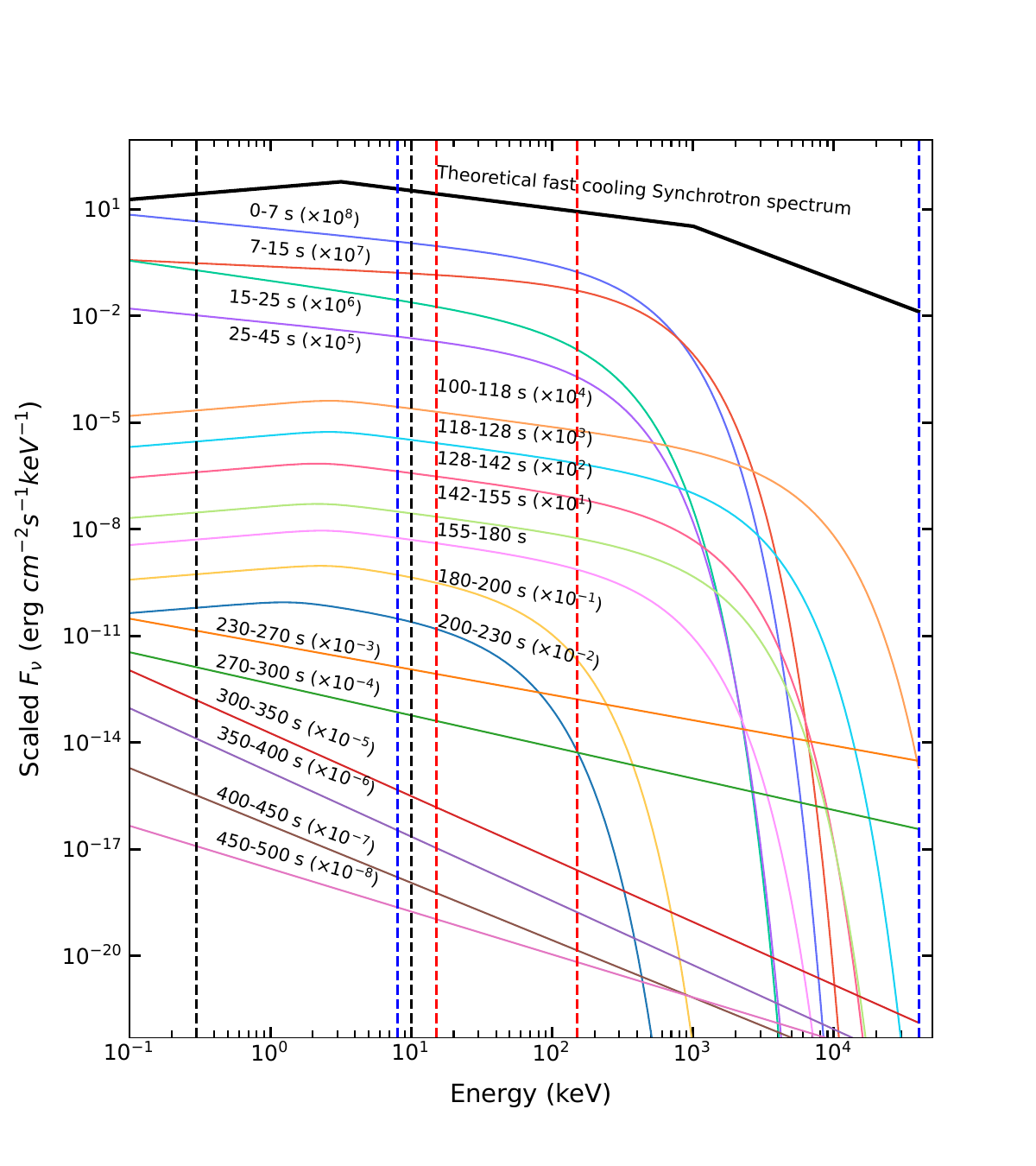}
 \caption{\textbf{Right}: Spectral energy distributions (SEDs) using parameters values of best joint spectral fitting in the coarse time bins. Four time slices are used in the First episode and seven time slices are obtained in the second episode. We also include 6 SEDs from late times after the second episode (from 230 to 500 sec). \textbf{Left}: Same SEDs with photon indices fixed at theoretically expected values ($\alpha_1$=-0.67 and $\alpha_2$=-1.5). }
 \label{fig:sed}
\end{figure*}

\section{Discussion}

% \subsection{Observed spectral signatures and synchrotron consistency}
\subsection{Synchrotron signatures in the time-resolved spectra}

The time-resolved spectra presented above already reveal synchrotron-like signatures in the prompt emission of GRB~241030A. In the early part of the second episode (100–230~s), the low-energy photon index $\alpha_1$ fluctuates around the fast-cooling synchrotron prediction of $-2/3$, while the higher-energy index $\alpha_2$ clusters near $-3/2$. During the later portion of the second episode (230–500~s), when the source is detected only by \xrt, the spectrum evolves from a relatively hard slope of $\sim\!-1.8$ toward a much softer index of $\sim\!-2.7$, broadly consistent with a transition from the fast-cooling value $-3/2$ to the asymptotic $-(p+2)/2$ segment expected when the cooling frequency drops below the observed band. Figure~\ref{fig:sed} compares the observed broadband SEDs with the theoretically expected synchrotron segments \citep{1998ApJ...497L..17S}. The overall agreement between the data and the predicted slopes indicates that the prompt emission can be explained naturally within the synchrotron framework.

In \S 5.2, we consider two widely discussed scenarios that could, in principle, account for these spectral properties: (i) a globally magnetically dominated outflow with a decaying field \citep{2011ApJ...726...90Z}, and (ii) the standard internal-shock framework. We find that the internal-shock scenario provides a more natural and flexible explanation of both the spectral shapes and their temporal evolution.

\subsection{Discriminating prompt–emission mechanisms: magnetized outflows versus internal shocks}
\label{sec:scenarios}

We first summarize the spectral expectations for a globally magnetized outflow with a decaying comoving field and compare them with the data. We then develop the internal–shock interpretation, deriving the relevant parameter scalings and confronting them with the observed evolution of $E_b$ and $E_p$. Next we use the near–zero spectral lag to place timing constraints that favor internal shocks. Finally, we discuss implications for the microphysics and jet composition.

% \subsection{Comparison of emission scenarios}
% \subsubsection{spectral properties in the magnetic dominated scenarios}
\subsubsection{Spectral expectations in magnetically dominated outflows}

One widely discussed scenario for GRB prompt emission posits a globally magnetized outflow \citep[e.g.,][]{2011ApJ...726...90Z,2014NatPh..10..351U}, in which the comoving magnetic field $B'$ decays as the ejecta expand. In the synchrotron framework, the spectral peak frequency $\nu_m$ (corresponding to $E_p$) in a time bin is
\begin{equation}
\nu_m \simeq \frac{3 q_e}{4\pi m_e c}\,\Gamma\,B_{\rm eff}\,\gamma_m^2,
\end{equation}
where $B_{\rm eff}$ is the effective magnetic field within the bin, $\gamma_m$ is the minimum Lorentz factor of the accelerated electrons, and $\Gamma$ is the bulk Lorentz factor. The synchrotron cooling frequency $\nu_c$ (corresponding to $E_b$) scales as
\begin{equation}
\nu_c \propto \Gamma\,B_{\rm eff}^{-3}\,(\Delta t')^{-2},
\end{equation}
with $\Delta t'$ the effective comoving cooling timescale for that bin.

Because the comoving field $B'$ decays with time, $\nu_c$ is expected to \emph{increase} systematically across bins. This is inconsistent with the approximately constant $E_b$ revealed by our time-resolved analysis. Moreover, a decaying magnetic field typically drives hard-to-soft $E_p$ evolution \citep[][and references therein]{2011ApJ...726...90Z} rather than the clear hardness–intensity tracking we observe. Magnetically dominated outflows also tend to produce appreciable spectral lags due to large emission radii and field decay \citep[e.g.,][]{2016ApJ...825...97U}, contrary to the nearly zero lags we measure. We therefore conclude that a globally magnetized outflow with a decaying comoving field faces significant difficulties in explaining the prompt emission of GRB~241030A.

% \subsubsection{Spectral Interpretation within the Internal Shock Scenario}

\subsubsection{Internal-shock interpretation and parameter scalings}
\label{subsec:internalshock}

Given the limitations of the globally magnetized outflow model, we consider the standard internal-shock scenario as an alternative explanation for the prompt emission of this burst. In this picture, two shells with different velocities collide at a radius $R$ and drive internal shocks. The relative Lorentz factor is
\[
\gamma_{\rm rel} \approx \frac{1}{2}\!\left(\frac{\gamma_f}{\gamma_s}+\frac{\gamma_s}{\gamma_f}\right),
\]
where $\gamma_f$ and $\gamma_s$ are the Lorentz factors of the fast and slow shells, respectively. The comoving proton number density, internal energy density, and magnetic-field strength behind the shock are respectively \citep{sariHydrodynamicTimescalesTemporal1995}
\begin{align}
n'_p &\simeq \frac{L}{4\pi R^{2}\Gamma^{2} m_{p}c^{3}},\\
e' &\simeq 4\,\gamma_{\rm rel}(\gamma_{\rm rel}-1)\,n'_p m_{p}c^{2},\\
B' &= \sqrt{8\pi\epsilon_{B}e'}
 \;\simeq\;
 \left[\frac{8\,\epsilon_{B}\,\gamma_{\rm rel}(\gamma_{\rm rel}-1)\,L}{R^{2}\Gamma^{2}c}\right]^{1/2},
\end{align}
where $L$ is the isotropic-equivalent kinetic luminosity of the outflow, $\Gamma$ is the bulk Lorentz factor of the shocked region in the observer frame, $m_p$ is the proton mass, $c$ is the speed of light, and $\epsilon_B$ is the fraction of post-shock internal energy in magnetic fields.

The magnetic field, the minimum electron Lorentz factor $\gamma_m$, and the synchrotron typical frequency $\nu_m$ scale as \citep{1998ApJ...497L..17S}
\begin{align}
B' &\propto R^{-1}\!\left[(\gamma_{\rm rel}-1)\gamma_{\rm rel}\right]^{1/2}\epsilon_B^{1/2},\\
\gamma_m &\propto (\gamma_{\rm rel}-1)\,(\epsilon_e/\xi),\\
\nu_m \propto B'\gamma_m^2 &\propto R^{-1}(\gamma_{\rm rel}-1)^{5/2}\gamma_{\rm rel}^{1/2}(\epsilon_e/\xi)^{2}\epsilon_B^{1/2},
\end{align}
where $\epsilon_e$ is the fraction of post-shock internal energy in electrons and $\xi$ is the fraction of electrons that are accelerated. The synchrotron cooling frequency scales as
\begin{equation}
\nu_c \propto \big(B'^3 \delta t^{2}\big)^{-1} \;\propto\; R^{3} \delta t^{-2}\left[(\gamma_{\rm rel}-1)\gamma_{\rm rel}\right]^{-3/2}\epsilon_B^{-3/2},
\end{equation}
with $\delta t$ being the (approximately fixed) time width of time-resolved spectra.

These scalings imply that variations in $R$ or $\epsilon_B$ drive $\nu_m$ and $\nu_c$ in opposite directions. For moderately relativistic internal shocks with large velocity contrast ($\gamma_{\rm rel}\gg 1$),
\[
\nu_m \propto \gamma_{\rm rel}^{3}, \qquad \nu_c \propto \gamma_{\rm rel}^{-3},
\]
so both break frequencies respond strongly—yet oppositely—to changes in $\gamma_{\rm rel}$. For mildly relativistic shocks ($\gamma_{\rm rel}\gtrsim 2$), the dependences weaken to
\[
\nu_m \propto \gamma_{\rm rel}^{1/2}, \qquad \nu_c \propto \gamma_{\rm rel}^{-3/2},
\]
but remain opposite in sign. Consequently, changes in $R$, $\epsilon_B$, or $\gamma_{\rm rel}$ alone would inevitably force $\nu_m$ and $\nu_c$ to evolve in opposite directions. This is inconsistent with our observations, which show that $\nu_m$ (i.e., $E_p$) exhibits a non-monotonic decrease with intermittent recoveries while $\nu_c$ (i.e., $E_b$) stays roughly constant (see Table~2). A natural resolution is that $\epsilon_e/\xi$ evolves with time while the other parameters remain approximately steady: since $\nu_m \propto (\epsilon_e/\xi)^2$ but $\nu_c$ is independent of $\epsilon_e$ and $\xi$, variations in $\epsilon_e/\xi$ can drive the observed evolution of $\nu_m$ without appreciably affecting $\nu_c$.

The overall time-resolved spectral properties of the two episodes can therefore be interpreted primarily through variations in $\gamma_m$ (and thus $\nu_m$). Table~1 shows that the spectral peak energies in Episode~I are generally lower than in Episode~II, consistent with a smaller $\gamma_m$ and hence a lower $\nu_m$. In Episode~I, the observed low-energy photon index of $-3/2$ suggests that $\nu_c$ lies below the GBM band ($\lesssim 8~\mathrm{keV}$), but the lack of simultaneous XRT coverage precludes a direct constraint on $B'$ relative to Episode~II. In the early part of Episode~II, the coexistence of photon indices $-3/2$ and $-2/3$ indicates that $\nu_c$ passes through the XRT band. Later in Episode~II, pronounced spectral evolution appears: $\gamma_m$ decreases, pushing $\nu_m$ below the XRT band, and, because the bins are relatively long, $\nu_c$ also drops below XRT. The late-time ($\sim$230--500~s) spectra then approach the expected fast-cooling high-energy slope with photon index $\sim -(p+2)/2$ ($\sim -2.7$). In Episode~I and the early Episode~II bins, this steep segment is not apparent, likely due to limited high-energy photon statistics. Similar behavior has been reported in other time-resolved GRB spectra \citep[e.g.,][]{kanekoCompleteSpectralCatalog2006,gruberFermiGBMGammaRay2014}. Overall, the synchrotron model within the internal-shock framework can account for the observed spectral evolution.

Similar spectral evolution has been reported in other GRBs where the synchrotron interpretation has been examined in detail. 
For instance, in GRB~160625B \citep{2018A&A...613A..16R}, \mfcgrb\ \citep{2023NatAs...7...67G}, and GRB~171010 \citep{ravasioEvidenceTwoSpectral2019}, 
$E_{\rm p}$ shows a general decreasing trend or flux-tracking behavior throughout the prompt phase, 
while the low-energy break $E_{\rm b}$ remains nearly constant. 
In contrast, some bursts such as GRB~160821 and GRB~180720 display more complex or independent evolution between $E_{\rm p}$ and $E_{\rm b}$ \citep{ravasioEvidenceTwoSpectral2019}. 
This diversity likely reflects differences in the temporal evolution of the underlying microphysical parameters among bursts.

\subsubsection{Spectral lags and timing constraints}

The internal–shock scenario can also accommodate the nearly zero spectral lag of this burst. In fact, within this framework the curvature effect plus synchrotron cooling alone often \emph{underpredict} the lags observed in many GRBs \citep{wuSpectralLagsGammaRay2000,shenSpectralLagsEnergy2005,uhmUnderstandingGRBPrompt2016}. In our case, however, the lag is essentially vanishing, which is compatible with injection–dominated timing. The synchrotron cooling timescale for typical parameters is

\begin{equation}
t_{\rm cool}\simeq 10^{-5}\ {\rm s}\,
\left(\frac{B}{10^{4}\ {\rm G}}\right)^{-3/2}
\left(\frac{h\nu}{100\ {\rm keV}}\right)^{-1/2}
\Gamma_{2.5}^{-1/2}.
\end{equation}
Our time-resolved spectral analysis finds a cooling break at $h\nu_c\sim 2$~keV, which—if one identifies the bin width $\delta t$ with the synchrotron cooling time—implies
\begin{equation}
B \simeq 14\ {\rm G}\,
\Gamma_{2.5}^{-1/3}
\left(\frac{h\nu_c}{2\ {\rm keV}}\right)^{-1/3}
\left(\frac{\delta t}{2.5\ {\rm s}}\right)^{-2/3},
\end{equation}
and hence
\begin{equation}
t_{\rm cool}\simeq 0.3\ {\rm s}\,
\left(\frac{B}{14\ {\rm G}}\right)^{-3/2}
\left(\frac{h\nu}{100\ {\rm keV}}\right)^{-1/2}
\Gamma_{2.5}^{-1/2}.
\end{equation}
A seconds-level cooling time is difficult to reconcile with the nearly zero spectral lag.

This tension arises from the simplifying assumption that the \emph{time-bin width equals the cooling time}. If a bin contains many short pulses, the relevant cooling time for each pulse is set by the \emph{local dynamical time} of the emitting region (e.g., the reverse-shock crossing time in a shell collision; \citealt{kobayashiCanInternalShocks1997}). Beyond this timescale, magnetic turbulence damps and/or particles are advected out of the high–field zone, radiative cooling no longer dominates, and the observed cooling break effectively “freezes’’ at the value reached when injection ceases. The spectrum fitted in a given bin is therefore a luminosity–weighted average over many pulses. Using the bin duration as the cooling time thus overestimates the cooling timescale and underestimates $B$.

Adopting the MVT $\sim 0.1$~s derived in Section~3.1 as a proxy for the cooling time of a typical short pulse yields
$ B \sim 1.2\times 10^{2}\ {\rm G}$ and $t_{\rm cool}\sim 10^{-2}\ {\rm s}$,
which is consistent with the observed near–zero lag. We note that even this field is smaller than the typical equipartition value expected in internal shocks \citep{piranGammarayBurstsFireball1999}. Moreover, the MVT is brightness–dependent and provides only an upper limit; a smaller intrinsic MVT would imply a larger $B$, further alleviating the tension. Thus, the internal–shock model can self–consistently account for both the timing and spectral properties.

\subsubsection{Microphysics and jet composition}

The synchrotron framework provides a useful way to probe how key microphysical parameters in internal shocks evolve during the burst. As discussed above, changes in $E_{\rm p}$ mainly trace variations in $\epsilon_e/\xi$. Previous studies suggest that $\epsilon_e$ is typically of order $0.1$ and does not vary strongly within individual bursts \citep[e.g.,][]{panaitescuPropertiesRelativisticJets2002,yostStudyAfterglowsFour2003}; therefore, the observed $E_{\rm p}$ evolution is likely governed by $\xi$. During $100$--$230$~s, $E_{\rm p}$ closely tracks the flux, implying an anticorrelation between $\xi$ and pulse intensity: stronger pulses would accelerate a smaller fraction of electrons. This trend is counterintuitive, since stronger shocks might be expected to energize more electrons. Particle-in-cell (PIC) simulations show that, in highly relativistic shocks, the fraction of electrons injected into the nonthermal tail depends sensitively on local shock conditions \citep[e.g.,][]{sironiPARTICLEACCELERATIONRELATIVISTIC2011,sironiMaximumEnergyAccelerated2013}, but the physical origin of a possible anticorrelation between shock strength and $\xi$ remains unclear. Moreover, the assumption that $\xi$ alone varies while the magnetic field stays nearly constant lacks firm physical justification. Some analytical GRB models posit that more violent shocks accelerate a larger fraction of electrons 

The spectral behavior also informs the jet composition. The consistency of the prompt spectra with synchrotron emission produced in internal shocks suggests that the outflow in this burst is likely matter dominated. In this picture, the prompt emission is primarily generated by collisions between unsteady baryonic shells, rather than by large-scale magnetic reconnection in a highly magnetized jet \citep[e.g.,][]{ZhangYan2011}. This highlights the diversity of energy-dissipation mechanisms in GRBs and indicates that, at least for some bursts, kinetic–energy–dominated outflows can power the prompt radiation.

\section{Summary and Conclusions}
\label{sec:sum}

We performed a joint, time–resolved \textit{Swift}–\textit{Fermi} analysis of the prompt emission of \mwgrb\ and showed that its spectra are well described by a smoothly connected broken power law over broad energy ranges. The burst comprises two distinct episodes. Episode~I (0–45~s) exhibits a fast–cooling synchrotron slope ($\alpha\!\sim\!-3/2$). In the early part of Episode~II (100–200~s), the spectra display both the $\alpha_1\!\simeq\!-2/3$ and $\alpha_2\!\simeq\!-3/2$ segments as the cooling break passes through the soft X–ray band, while at later times ($\gtrsim$230~s) the spectrum softens toward $\sim\!-2.7$. Throughout Episode~II, the low–energy break remains nearly constant at a few keV (naturally identified with $\nu_c$), whereas the spectral peak $E_{\rm p}$ tracks the flux within individual pulses and steps down between them. The spectral lag across GBM bands is consistent with zero.

These combined properties are challenging for a globally magnetized outflow with a decaying comoving field, which typically predicts a growing $\nu_c$, hard–to–soft $E_{\rm p}$ evolution, and appreciable lags. By contrast, they arise naturally in an internal–shock synchrotron scenario in which the effective magnetic field is roughly steady while the fraction of accelerated electrons (equivalently, the minimum electron Lorentz factor) varies in time: this reproduces the stable $E_b$, the intensity–tracking yet step–down $E_{\rm p}$, the canonical $-2/3$ and $-3/2$ slopes, and the near–zero lag (especially when the relevant cooling time is set by the minimum variability timescale rather than by the bin width).

We also compared single–component fits with a two–component (BB+CPL) prescription and found statistically comparable descriptions of the time–resolved spectra. In the absence of independent evidence for a thermal component, we find no compelling need to introduce an additional spectral component for \mwgrb.

Overall, our results favor a baryonic, matter–dominated jet in which the prompt radiation of \mwgrb\ is produced by fast–cooling synchrotron emission from internal shocks, rather than by magnetic dissipation in a highly magnetized outflow.

\begin{table*}
\centering
\caption{Coarse time binning spectral fitting data}
\label{tab:cbin}
\begin{tabular}{|c|c|c|c|c|c|c|c|c|c|}
\toprule
 T1 & T2 & $\alpha_1$ & $\alpha_2$ & $E_b$ & $E_p$ & logA & Statistic/DOF & AIC & BIC \\
\midrule
\hline
\multicolumn{10}{|c|}{Model: CPL}\\
\hline
 0 & 7 & & $-1.39_{-0.28}^{+0.12}$ & & $108.12_{-0.38}^{+2.65}$ & $-2.52_{-0.22}^{+0.09}$ & 112/110 & 116 & 124 \\
 7 & 15 & & $-1.18_{-0.08}^{+0.07}$ & & $188.10_{-0.33}^{+0.38}$ & $-2.17_{-0.05}^{+0.05}$ & 130/150 & 136 & 145 \\
 15 & 25 & & $-1.58_{-0.16}^{+0.14}$ & & $39.81_{-0.23}^{+0.17}$ & $-2.35_{-0.15}^{+0.12}$ & 144/135 & 149 & 158 \\
 25 & 45 & & $-1.39_{-0.09}^{+0.08}$ & & $59.78_{-0.11}^{+0.11}$ & $-2.18_{-0.07}^{+0.07}$ & 174/172 & 179 & 189 \\
 \hline
 0 & 45 & & $-1.43_{-0.08}^{+0.06}$ & & $77.62_{-1.28}^{+1.14}$ & $-2.29_{-0.06}^{+0.05}$ & 208/172 & 214 & 223 \\ 
 \hline
 \multicolumn{10}{|c|}{Model: CSBPL}\\
 \hline
100 & 118 & $-0.40_{-0.10}^{+0.10}$ & $-1.57_{-0.02}^{+0.01}$ & $2.87_{-0.18}^{+1.79}$ & $868.76_{-0.54}^{+1.83}$ & $0.32_{-0.03}^{+0.02}$ & 450/352 & 460 & 480 \\
118 & 128 & $-0.22_{-0.12}^{+0.15}$ & $-1.60_{-0.03}^{+0.02}$ & $2.84_{-0.20}^{+0.74}$ & $470.47_{-0.66}^{+0.67}$ & $0.43_{-0.04}^{+0.03}$ & 411/308 & 421 & 439 \\
128 & 142 & $-0.37_{-0.13}^{+0.13}$ & $-1.36_{-0.02}^{+0.01}$ & $2.30_{-0.20}^{+0.16}$ & $240.60_{-0.13}^{+0.16}$ & $0.58_{-0.03}^{+0.03}$ & 649/371 & 659 & 679 \\
142 & 155 & $-0.28_{-0.12}^{+0.15}$ & $-1.28_{-0.01}^{+0.02}$ & $2.57_{-0.35}^{+0.12}$ & $278.18_{-0.13}^{+0.11}$ & $0.45_{-0.03}^{+0.03}$ & 714/385 & 724 & 743 \\
155 & 180 & $-0.47_{-0.09}^{+0.09}$ & $-1.50_{-0.02}^{+0.02}$ & $2.58_{-0.24}^{+0.15}$ & $120.61_{-0.13}^{+0.12}$ & $0.68_{-0.01}^{+0.03}$ & 818/416 & 828 & 848 \\
180 & 200 & $-0.69_{-0.07}^{+0.10}$ & $-1.98_{-0.01}^{+0.02}$ & $2.71_{-0.23}^{+9.01}$ & $21.05_{-0.13}^{+4.39}$ & $0.70_{-0.02}^{+0.02}$ & 354/317 & 364 & 383 \\
200 & 230 & $-0.85_{-0.09}^{+0.15}$ & $-1.84_{-0.02}^{+0.04}$ & $1.44_{-0.14}^{+0.33}$ & $10.01_{-0.01}^{+0.02}$ & $0.77_{-0.01}^{+0.03}$ & 470/326 & 480 & 499 \\
\hline
100 & 230 & $-0.21_{-0.03}^{+1.21}$ & $-1.59_{-0.01}^{+0.16}$ & $ 1.81_{-1.47}^{+1.04}$ & $ 281.84_{-1.65}^{+1.07}$ & $0.66_{-0.02}^{+0.06}$ & 1594/781 & 1604 & 1627 \\
\hline
\multicolumn{10}{|c|}{Model: PL}\\
\hline
230 & 270 & $-1.71_{-0.02}^{+0.02}$ & & & & & 290/230 & 294 & 300 \\
270 & 300 & $-1.88_{-0.03}^{+0.03}$ & & & & & 169/152 & 163 & 169 \\
300 & 350 & $-2.77_{-0.03}^{+0.03}$ & & & & & 223/177 & 227 & 233 \\
350 & 400 & $-2.80_{-0.02}^{+0.03}$ & & & & & 215/188 & 219 & 225 \\
400 & 450 & $-2.61_{-0.05}^{+0.05}$ & & & & & 153/100 & 157 & 162 \\ 
450 & 500 & $-2.20_{-0.05}^{+0.06}$ & & & & & 103/72 & 107 & 111 \\
\bottomrule
\end{tabular}
\end{table*}

\begin{table*}
\centering
\caption{Fine time binning spectral fitting data.}
\label{tab:spec_fit1}
\begin{tabular}{|c|c|c|c|c|c|c|c|c|}
\hline
\hline
\multicolumn{9}{c}{Joint fit} \\
\hline
 & & \multicolumn{7}{c|}{csbpl}\\
 \hline
 t1 (s) & t2 (s) & $\alpha_1$ & $\alpha_2$ & $E_b$ & $E_p$ & logA & stat/dof & BIC \\
\hline
 104.5 & 107.0 & $-0.77_{-0.2}^{+0.24}$ & $-1.98_{-0.0}^{+0.14}$ & $3.63_{-0.13}^{+0.10}$ & $30.20_{-0.61}^{+2.22}$ & $0.35_{-0.08}^{+0.05}$ & 97.95/98.0 & 120.99 \\
 107.0 & 109.5 & $-0.45_{-0.21}^{+0.43}$ & $-1.78_{-0.03}^{+0.09}$ & $2.95_{-0.13}^{+0.09}$ & $870.96_{-4.74}^{+0.65}$ & $0.36_{-0.11}^{+0.05}$ & 121.44/104.0 & 144.90 \\
 109.5 & 112.0 & $-0.53_{-0.19}^{+0.24}$ & $-1.36_{-0.02}^{+0.06}$ & $4.17_{-0.19}^{+0.17}$ & $575.44_{-0.89}^{+0.14}$ & $0.27_{-0.07}^{+0.06}$ & 140.55/157.0 & 165.99 \\
 112.0 & 114.5 & $1.15_{-0.33}^{+0.53}$ & $-1.42_{-0.04}^{+0.02}$ & $2.45_{-0.03}^{+0.04}$ & $467.74_{-0.37}^{+0.13}$ & $-0.02_{-0.14}^{+0.08}$ & 227.68/194.0 & 254.14 \\
 114.5 & 117.0 & $0.45_{-0.47}^{+0.52}$ & $-1.42_{-0.03}^{+0.02}$ & $2.19_{-0.04}^{+0.05}$ & $549.54_{-0.63}^{+0.09}$ & $0.28_{-0.09}^{+0.11}$ & 244.52/178.0 & 270.57 \\
 117.0 & 119.5 & $-0.19_{-0.12}^{+0.58}$ & $-1.98_{-0.01}^{+0.05}$ & $2.69_{-0.06}^{+0.03}$ & $18.62_{-0.03}^{+2.00}$ & $0.38_{-0.16}^{+0.01}$ & 117.71/103.0 & 140.73 \\
 119.5 & 122.0 & $-0.25_{-0.12}^{+0.4}$ & $-1.7_{-0.08}^{+0.08}$ & $3.72_{-0.12}^{+0.05}$ & $66.07_{-0.21}^{+0.20}$ & $0.36_{-0.10}^{+0.05}$ & 134.68/116.0 & 158.66 \\
 122.0 & 124.5 & $-0.1_{-0.19}^{+0.55}$ & $-1.27_{-0.02}^{+0.04}$ & $2.34_{-0.09}^{+0.03}$ & $389.05_{-0.48}^{+0.02}$ & $0.46_{-0.09}^{+0.05}$ & 277.46/217.0 & 304.47 \\
 124.5 & 127.0 & $-0.34_{-0.14}^{+0.29}$ & $-1.92_{-0.04}^{+0.05}$ & $3.24_{-0.08}^{+0.05}$ & $2630.27_{-7.24}^{+0.40}$ & $0.49_{-0.08}^{+0.04}$ & 127.6/120.0 & 151.38 \\
 127.0 & 129.5 & $-0.79_{-0.09}^{+0.19}$ & $-1.74_{-0.15}^{+0.06}$ & $5.25_{-0.15}^{+0.17}$ & $97.72_{-0.69}^{+1.46}$ & $0.55_{-0.06}^{+0.04}$ & 136.86/129.0 & 161.35 \\
 129.5 & 132.0 & $-0.4_{-0.34}^{+0.44}$ & $-1.33_{-0.03}^{+0.08}$ & $1.91_{-0.14}^{+0.06}$ & $107.15_{-0.23}^{+0.03}$ & $0.61_{-0.03}^{+0.12}$ & 151.25/148.0 & 176.41 \\
 132.0 & 134.5 & $-0.59_{-0.1}^{+0.19}$ & $-1.43_{-0.02}^{+0.07}$ & $4.57_{-0.21}^{+0.09}$ & $190.55_{-0.31}^{+0.09}$ & $0.62_{-0.06}^{+0.04}$ & 277.73/208.0 & 304.53 \\
 134.5 & 137.0 & $-0.28_{-0.18}^{+0.56}$ & $-1.37_{-0.01}^{+0.05}$ & $2.24_{-0.09}^{+0.04}$ & $346.74_{-0.58}^{+0.04}$ & $0.55_{-0.08}^{+0.05}$ & 263.8/193.0 & 290.24 \\
 137.0 & 139.5 & $-0.32_{-0.35}^{+0.81}$ & $-1.18_{-0.02}^{+0.05}$ & $1.78_{-0.26}^{+0.03}$ & $407.38_{-0.24}^{+0.03}$ & $0.51_{-0.06}^{+0.32}$ & 251.24/220.0 & 278.32 \\
 139.5 & 142.0 & $-0.48_{-0.29}^{+1.04}$ & $-1.35_{-0.01}^{+0.07}$ & $1.48_{-0.20}^{+0.02}$ & $398.11_{-0.66}^{+0.03}$ & $0.58_{-0.04}^{+0.54}$ & 207.12/159.0 & 232.62 \\
 142.0 & 144.5 & $0.16_{-0.8}^{+0.93}$ & $-1.13_{-0.03}^{+0.04}$ & $1.10_{-0.04}^{+0.02}$ & $239.88_{-0.22}^{+0.00}$ & $0.54_{-0.05}^{+0.73}$ & 198.98/187.0 & 225.27 \\
 144.5 & 147.0 & $-0.58_{-0.37}^{+0.37}$ & $-1.08_{-0.02}^{+0.05}$ & $2.34_{-0.34}^{+0.13}$ & $331.13_{-0.29}^{+0.02}$ & $0.45_{-0.02}^{+0.15}$ & 258.81/225.0 & 286.00 \\
 147.0 & 149.5 & $-0.27_{-0.24}^{+0.26}$ & $-1.01_{-0.01}^{+0.04}$ & $2.75_{-0.10}^{+0.12}$ & $263.03_{-0.17}^{+0.01}$ & $0.41_{-0.07}^{+0.07}$ & 359.59/237.0 & 387.04 \\
 149.5 & 152.0 & $-0.61_{-0.08}^{+0.16}$ & $-1.78_{-0.06}^{+0.05}$ & $6.92_{-0.14}^{+0.08}$ & $194.98_{-0.95}^{+0.58}$ & $0.50_{-0.06}^{+0.05}$ & 233.31/162.0 & 258.90 \\
 152.0 & 154.5 & $0.08_{-0.37}^{+0.53}$ & $-1.61_{-0.11}^{+0.07}$ & $2.63_{-0.07}^{+0.08}$ & $33.11_{-0.28}^{+0.04}$ & $0.42_{-0.11}^{+0.08}$ & 162.81/130.0 & 187.34 \\
 154.5 & 157.0 & $-0.65_{-0.1}^{+0.37}$ & $-1.61_{-0.16}^{+0.08}$ & $3.39_{-0.13}^{+0.15}$ & $25.12_{-0.42}^{+0.05}$ & $0.65_{-0.08}^{+0.03}$ & 97.41/117.0 & 121.43 \\
 157.0 & 159.5 & $-0.5_{-0.11}^{+0.17}$ & $-1.88_{-0.06}^{+0.09}$ & $4.90_{-0.13}^{+0.06}$ & $26.92_{-0.33}^{+0.24}$ & $0.71_{-0.04}^{+0.05}$ & 169.39/144.0 & 194.41 \\
 159.5 & 162.0 & $0.99_{-0.69}^{+0.27}$ & $-1.27_{-0.02}^{+0.03}$ & $1.74_{-0.01}^{+0.04}$ & $177.83_{-0.16}^{+0.02}$ & $0.54_{-0.08}^{+0.09}$ & 279.07/219.0 & 306.13 \\
 162.0 & 164.5 & $-0.65_{-0.11}^{+0.91}$ & $-1.15_{-0.01}^{+0.04}$ & $1.32_{-0.21}^{+0.01}$ & $144.54_{-0.15}^{+0.01}$ & $0.83_{-0.03}^{+0.55}$ & 351.47/213.0 & 378.40 \\
 164.5 & 167.0 & $-0.57_{-0.17}^{+0.19}$ & $-1.48_{-0.04}^{+0.1}$ & $3.31_{-0.13}^{+0.07}$ & $70.79_{-0.21}^{+0.04}$ & $0.74_{-0.05}^{+0.06}$ & 183.51/169.0 & 209.30 \\
 167.0 & 169.5 & $-0.34_{-0.2}^{+0.59}$ & $-1.24_{-0.05}^{+0.05}$ & $1.86_{-0.09}^{+0.06}$ & $97.72_{-0.14}^{+0.02}$ & $0.73_{-0.05}^{+0.08}$ & 209.47/189.0 & 235.81 \\
 169.5 & 172.0 & $-0.68_{-0.17}^{+0.2}$ & $-1.59_{-0.1}^{+0.06}$ & $2.75_{-0.07}^{+0.15}$ & $53.70_{-0.16}^{+0.06}$ & $0.81_{-0.08}^{+0.03}$ & 207.35/151.0 & 232.60 \\
 172.0 & 174.5 & $-0.05_{-0.12}^{+1.3}$ & $-1.91_{-0.08}^{+0.06}$ & $1.48_{-0.04}^{+0.01}$ & $11.22_{-0.05}^{+0.35}$ & $0.68_{-0.09}^{+0.05}$ & 136.18/103.0 & 159.10 \\
 174.5 & 177.0 & $0.47_{-1.02}^{+0.51}$ & $-1.8_{-0.04}^{+0.18}$ & $1.48_{-0.24}^{+0.02}$ & $10.23_{-0.00}^{+0.02}$ & $1.22_{-0.24}^{+1.11}$ & 147.23/102.0 & 169.69 \\
 177.0 & 179.5 & $0.61_{-1.13}^{+0.68}$ & $-1.65_{-0.23}^{+0.12}$ & $1.12_{-0.02}^{+0.01}$ & $10.23_{-0.00}^{+0.01}$ & $0.86_{-0.14}^{+0.20}$ & 145.36/100.0 & 167.99 \\
 179.5 & 182.0 & $-0.89_{-0.42}^{+0.41}$ & $-2.0_{-0.0}^{+0.05}$ & $1.32_{-0.11}^{+0.00}$ & $24.55_{-0.77}^{+0.47}$ & $0.67_{-0.00}^{+0.30}$ & 163.33/100.0 & 186.27 \\
 182.0 & 184.5 & $-0.72_{-0.31}^{+0.94}$ & $-1.81_{-0.06}^{+0.12}$ & $1.58_{-0.08}^{+0.03}$ & $11.22_{-0.00}^{+0.10}$ & $0.69_{-0.07}^{+0.07}$ & 137.02/99.0 & 159.42 \\
 184.5 & 187.0 & $-0.8_{-0.15}^{+0.22}$ & $-1.91_{-0.03}^{+0.11}$ & $3.55_{-0.15}^{+0.03}$ & $14.45_{-0.08}^{+0.34}$ & $0.70_{-0.05}^{+0.06}$ & 158.92/105.0 & 182.38 \\
 187.0 & 189.5 & $-0.28_{-0.28}^{+0.51}$ & $-1.42_{-0.06}^{+0.03}$ & $1.95_{-0.03}^{+0.08}$ & $134.90_{-0.25}^{+0.05}$ & $0.74_{-0.10}^{+0.03}$ & 211.66/179.0 & 237.74 \\
 189.5 & 192.0 & $-0.63_{-0.16}^{+0.22}$ & $-1.88_{-0.01}^{+0.1}$ & $3.24_{-0.09}^{+0.05}$ & $10.72_{-0.02}^{+0.14}$ & $0.70_{-0.06}^{+0.05}$ & 129.82/112.0 & 153.45 \\
 192.0 & 194.5 & $-0.29_{-0.37}^{+1.0}$ & $-1.54_{-0.09}^{+0.2}$ & $1.15_{-0.05}^{+0.01}$ & $10.00_{-0.00}^{+0.00}$ & $0.85_{-0.05}^{+0.31}$ & 112.73/103.0 & 136.14 \\
 104.5 & 194.5 & $-0.49_{-0.05}^{+0.05}$ & $-1.49_{-0.01}^{+0.02}$ & $3.02_{-0.03}^{+0.02}$ & $251.19_{-0.17}^{+0.04}$ & $0.52_{-0.01}^{+0.06}$ & 639.12/478.0 & 670.02 \\
\hline
\hline
\end{tabular}
\end{table*}

\begin{table*}
\centering
\small
\caption{Fine time binning fitting results.}
\label{tab:spec_fit2}
\begin{tabular}{|c|c|c|c|c|c|c|c|c|}
\hline
\hline
\multicolumn{9}{c|}{Joint fit}\\
\hline
\multirow{2}{*}{t1 (s)} & \multirow{2}{*}{t2 (s)} & \multicolumn{7}{c}{bb+cpl}\\
\cline{3-9}
& & log$kT$ & log$A_{bb}$ & $\alpha$ & log$E_{p}$ & log$A_{cpl}$ & STAT/dof & BIC\\
\hline
104.50 & 107.00 & ${0.18}_{-0.04}^{+0.13}$ & ${-0.69}_{-0.17}^{+0.12}$ & ${-1.34}_{-0.16}^{+0.01}$ & ${1.63}_{-0.09}^{+0.22}$ & ${-2.41}_{-0.23}^{+0.07}$ & 96.31/98 & 119.49\\
107.00 & 109.50 & ${0.15}_{-0.06}^{+0.05}$ & ${-0.52}_{-0.15}^{+0.04}$ & ${-1.20}_{-0.10}^{+0.00}$ & ${1.95}_{-0.02}^{+0.14}$ & ${-2.17}_{-0.10}^{+0.00}$ & 114.92/104 & 138.38\\
109.50 & 112.00 & ${0.51}_{-0.05}^{+0.14}$ & ${-0.45}_{-0.20}^{+0.03}$ & ${-1.10}_{-0.05}^{+0.02}$ & ${2.52}_{-0.07}^{+0.08}$ & ${-1.84}_{-0.04}^{+0.02}$ & 145.03/157 & 170.47\\
112.00 & 114.50 & ${0.41}_{-0.02}^{+0.05}$ & ${-0.11}_{-0.07}^{+0.03}$ & ${-1.00}_{-0.04}^{+0.03}$ & ${2.36}_{-0.04}^{+0.05}$ & ${-1.68}_{-0.03}^{+0.02}$ & 267.31/194 & 293.78\\
114.50 & 117.00 & ${0.32}_{-0.06}^{+0.05}$ & ${-0.34}_{-0.15}^{+0.06}$ & ${-1.15}_{-0.03}^{+0.04}$ & ${2.43}_{-0.06}^{+0.06}$ & ${-1.80}_{-0.03}^{+0.03}$ & 266.00/178 & 292.05\\
117.00 & 119.50 & ${0.12}_{-0.03}^{+0.05}$ & ${-0.37}_{-0.06}^{+0.06}$ & ${-1.16}_{-0.12}^{+0.17}$ & ${1.76}_{-0.07}^{+0.09}$ & ${-2.21}_{-0.12}^{+0.16}$ & 104.72/103 & 128.14\\
119.50 & 122.00 & ${0.25}_{-0.04}^{+0.03}$ & ${-0.20}_{-0.11}^{+0.04}$ & ${-1.00}_{-0.07}^{+0.08}$ & ${1.82}_{-0.04}^{+0.04}$ & ${-1.74}_{-0.07}^{+0.08}$ & 130.80/116 & 154.77\\
122.00 & 124.50 & ${0.42}_{-0.05}^{+0.03}$ & ${-0.12}_{-0.08}^{+0.05}$ & ${-1.03}_{-0.03}^{+0.02}$ & ${2.43}_{-0.03}^{+0.04}$ & ${-1.54}_{-0.02}^{+0.01}$ & 265.89/217 & 292.9\\
124.50 & 127.00 & ${0.23}_{-0.05}^{+0.03}$ & ${-0.19}_{-0.11}^{+0.02}$ & ${-1.30}_{-0.09}^{+0.04}$ & ${1.99}_{-0.08}^{+0.20}$ & ${-2.16}_{-0.11}^{+0.06}$ & 132.97/120 & 157.12\\
127.00 & 129.50 & ${0.33}_{-0.06}^{+0.04}$ & ${-0.37}_{-0.15}^{+0.08}$ & ${-1.17}_{-0.07}^{+0.04}$ & ${1.81}_{-0.05}^{+0.05}$ & ${-1.82}_{-0.07}^{+0.05}$ & 141.70/129 & 166.19\\
129.50 & 132.00 & ${0.09}_{-0.04}^{+0.20}$ & ${-0.58}_{-0.68}^{+0.06}$ & ${-1.09}_{-0.10}^{+0.03}$ & ${1.97}_{-0.02}^{+0.04}$ & ${-1.64}_{-0.07}^{+0.03}$ & 144.91/148 & 170.06\\
132.00 & 134.50 & ${0.47}_{-0.03}^{+0.04}$ & ${-0.10}_{-0.08}^{+0.05}$ & ${-1.09}_{-0.02}^{+0.03}$ & ${2.17}_{-0.04}^{+0.02}$ & ${-1.51}_{-0.02}^{+0.03}$ & 282.08/208 & 308.89\\
134.50 & 137.00 & ${0.29}_{-0.09}^{+0.05}$ & ${-0.33}_{-0.20}^{+0.03}$ & ${-1.13}_{-0.04}^{+0.03}$ & ${2.34}_{-0.05}^{+0.04}$ & ${-1.68}_{-0.03}^{+0.03}$ & 266.41/193 & 292.85\\
137.00 & 139.50 & ${0.34}_{-0.06}^{+1.30}$ & ${-0.48}_{-2.95}^{+0.05}$ & ${-1.08}_{-0.05}^{+0.01}$ & ${2.54}_{-0.02}^{+0.06}$ & ${-1.59}_{-0.03}^{+0.01}$ & 245.98/220 & 273.06\\
139.50 & 142.00 & ${0.09}_{-0.04}^{+0.06}$ & ${-0.53}_{-0.13}^{+0.06}$ & ${-1.12}_{-0.07}^{+0.03}$ & ${2.41}_{-0.06}^{+0.08}$ & ${-1.86}_{-0.04}^{+0.03}$ & 184.80/159 & 210.3\\
142.00 & 144.50 & ${2.40}_{-1.96}^{+0.14}$ & ${-4.34}_{-0.81}^{+2.95}$ & ${-1.10}_{-0.02}^{+0.02}$ & ${2.35}_{-0.03}^{+0.04}$ & ${-1.65}_{-0.02}^{+0.02}$ & 202.45/187 & 226.82\\
144.50 & 147.00 & ${0.35}_{-0.05}^{+1.99}$ & ${-0.61}_{-3.79}^{+0.07}$ & ${-0.99}_{-0.05}^{+0.02}$ & ${2.46}_{-0.02}^{+0.04}$ & ${-1.49}_{-0.02}^{+0.01}$ & 253.03/225 & 279.84\\
147.00 & 149.50 & ${0.86}_{-0.14}^{+0.08}$ & ${-0.36}_{-0.53}^{+0.03}$ & ${-0.89}_{-0.02}^{+0.02}$ & ${2.40}_{-0.04}^{+0.01}$ & ${-1.27}_{-0.00}^{+0.03}$ & 361.81/237 & 389.25\\
149.50 & 152.00 & ${0.54}_{-0.04}^{+0.06}$ & ${-0.10}_{-0.09}^{+0.02}$ & ${-1.17}_{-0.05}^{+0.04}$ & ${2.04}_{-0.06}^{+0.06}$ & ${-1.76}_{-0.05}^{+0.05}$ & 250.43/162 & 276.02\\
152.00 & 154.50 & ${0.27}_{-0.06}^{+0.04}$ & ${-0.22}_{-0.11}^{+0.05}$ & ${-1.01}_{-0.09}^{+0.06}$ & ${1.60}_{-0.03}^{+0.04}$ & ${-1.62}_{-0.10}^{+0.07}$ & 170.50/130 & 195.03\\
154.50 & 157.00 & ${0.16}_{-0.02}^{+0.10}$ & ${-0.40}_{-0.18}^{+0.06}$ & ${-1.11}_{-0.09}^{+0.00}$ & ${1.52}_{-0.03}^{+0.03}$ & ${-1.66}_{-0.11}^{+0.03}$ & 97.34/117 & 121.36\\
157.00 & 159.50 & ${1.49}_{-0.02}^{+0.07}$ & ${0.16}_{-0.05}^{+0.05}$ & ${-0.93}_{-0.06}^{+0.02}$ & ${1.31}_{-0.01}^{+0.04}$ & ${-1.01}_{-0.11}^{+0.01}$ & 184.66/144 & 209.46\\
159.50 & 162.00 & ${0.33}_{-0.03}^{+0.03}$ & ${0.03}_{-0.07}^{+0.06}$ & ${-0.97}_{-0.03}^{+0.04}$ & ${2.14}_{-0.02}^{+0.02}$ & ${-1.27}_{-0.02}^{+0.02}$ & 287.48/219 & 314.53\\
162.00 & 164.50 & ${0.21}_{-0.06}^{+0.03}$ & ${-0.26}_{-0.14}^{+0.10}$ & ${-1.02}_{-0.03}^{+0.04}$ & ${2.12}_{-0.02}^{+0.02}$ & ${-1.33}_{-0.02}^{+0.03}$ & 331.40/213 & 358.32\\
164.50 & 167.00 & ${0.19}_{-0.04}^{+0.05}$ & ${-0.24}_{-0.13}^{+0.08}$ & ${-1.08}_{-0.06}^{+0.04}$ & ${1.80}_{-0.02}^{+0.03}$ & ${-1.45}_{-0.05}^{+0.03}$ & 175.65/169 & 201.44\\
167.00 & 169.50 & ${0.22}_{-0.05}^{+0.04}$ & ${-0.18}_{-0.14}^{+0.06}$ & ${-1.02}_{-0.04}^{+0.05}$ & ${1.96}_{-0.03}^{+0.02}$ & ${-1.36}_{-0.03}^{+0.04}$ & 191.84/189 & 218.18\\
169.50 & 172.00 & ${0.14}_{-0.04}^{+0.06}$ & ${-0.30}_{-0.13}^{+0.07}$ & ${-1.19}_{-0.07}^{+0.02}$ & ${1.72}_{-0.02}^{+0.04}$ & ${-1.66}_{-0.07}^{+0.02}$ & 199.11/151 & 224.36\\
172.00 & 174.50 & ${0.02}_{-0.01}^{+0.07}$ & ${-0.42}_{-0.04}^{+0.08}$ & ${-1.50}_{-0.20}^{+0.02}$ & ${1.57}_{-0.07}^{+0.75}$ & ${-2.60}_{-0.33}^{+0.01}$ & 123.39/103 & 146.68\\
174.50 & 177.00 & ${0.93}_{-0.05}^{+0.10}$ & ${-0.75}_{-0.11}^{+0.05}$ & ${-0.39}_{-0.20}^{+0.48}$ & ${0.58}_{-0.05}^{+0.05}$ & ${0.16}_{-0.43}^{+1.00}$ & 110.46/102 & 133.83\\
177.00 & 179.50 & ${1.33}_{-0.62}^{+1.15}$ & ${-0.92}_{-3.78}^{+0.06}$ & ${-0.52}_{-0.27}^{+0.22}$ & ${0.73}_{-0.05}^{+0.05}$ & ${-0.26}_{-0.53}^{+0.44}$ & 102.44/100 & 125.71\\
179.50 & 182.00 & ${1.49}_{-0.11}^{+0.13}$ & ${-0.53}_{-0.22}^{+0.12}$ & ${-0.19}_{-0.55}^{+0.24}$ & ${0.54}_{-0.03}^{+0.11}$ & ${0.51}_{-1.18}^{+0.49}$ & 119.17/100 & 142.44\\
182.00 & 184.50 & ${0.07}_{-0.03}^{+0.15}$ & ${-0.69}_{-0.37}^{+0.02}$ & ${-1.53}_{-0.14}^{+0.07}$ & ${1.34}_{-0.16}^{+0.03}$ & ${-2.53}_{-0.17}^{+0.19}$ & 128.46/99 & 151.68\\
184.50 & 187.00 & ${0.13}_{-0.07}^{+0.03}$ & ${-0.37}_{-0.12}^{+0.07}$ & ${-1.34}_{-0.06}^{+0.09}$ & ${1.56}_{-0.04}^{+0.07}$ & ${-2.04}_{-0.09}^{+0.10}$ & 150.29/105 & 173.79\\
187.00 & 189.50 & ${0.31}_{-0.10}^{+0.07}$ & ${-0.36}_{-0.22}^{+0.09}$ & ${-1.20}_{-0.04}^{+0.04}$ & ${2.04}_{-0.04}^{+0.04}$ & ${-1.64}_{-0.03}^{+0.03}$ & 218.34/179 & 244.41\\
189.50 & 192.00 & ${0.19}_{-0.05}^{+0.05}$ & ${-0.17}_{-0.12}^{+0.03}$ & ${-1.28}_{-0.09}^{+0.08}$ & ${1.47}_{-0.06}^{+0.06}$ & ${-1.96}_{-0.13}^{+0.12}$ & 125.97/112 & 149.78\\
192.00 & 194.50 & ${0.80}_{-0.03}^{+0.28}$ & ${-0.59}_{-0.40}^{+0.02}$ & ${-0.44}_{-0.56}^{+0.06}$ & ${0.67}_{-0.01}^{+0.21}$ & ${0.02}_{-1.18}^{+0.12}$ & 98.15/103 & 121.56\\
104.50 & 194.50 & ${0.22}_{-0.01}^{+0.02}$ & ${-0.40}_{-0.02}^{+0.03}$ & ${-1.14}_{-0.01}^{+0.01}$ & ${2.17}_{-0.01}^{+0.02}$ & ${-1.79}_{-0.01}^{+0.01}$ & 658.69/478 & 689.59\\
\hline
\hline
\end{tabular}
\end{table*}

\begin{acknowledgments}

We acknowledge the support by the National Key Research and Development Programs of China (2022YFF0711404, 2022SKA0130102, 2021YFA0718500), the National SKA Program of China (2022SKA0130100), the National Natural Science Foundation of China (grant Nos. 11833003, U2038105, U1831135, 12121003, 12041301, 12393811, 13001106, 12403035, 12573046, 13001106), the science research grants from the China Manned Space Project with NO. CMS-CSST-2021-B11, the Fundamental Research Funds for the Central Universities.

\end{acknowledgments}

\bibliography{ms.bib}

\begin{thebibliography}{}
\expandafter\ifx\csname natexlab\endcsname\relax\def\natexlab#1{#1}\fi
\providecommand{\url}[1]{\href{#1}{#1}}
\providecommand{\dodoi}[1]{doi:~\href{http://doi.org/#1}{\nolinkurl{#1}}}
\providecommand{\doeprint}[1]{\href{http://ascl.net/#1}{\nolinkurl{http://ascl.net/#1}}}
\providecommand{\doarXiv}[1]{\href{https://arxiv.org/abs/#1}{\nolinkurl{https://arxiv.org/abs/#1}}}

\bibitem[{{Arnaud}(1996)}]{1996ASPC..101...17A}
{Arnaud}, K.~A. 1996, in Astronomical Society of the Pacific Conference Series, Vol. 101, Astronomical Data Analysis Software and Systems V, ed. G.~H. {Jacoby} \& J.~{Barnes}, 17

\bibitem[{{Beardmore} {et~al.}(2024){Beardmore}, {Evans}, {Goad}, {Osborne}, \& {Swift-XRT Team.}}]{2024GCN.37962....1B}
{Beardmore}, A.~P., {Evans}, P.~A., {Goad}, M.~R., {Osborne}, J.~P., \& {Swift-XRT Team.} 2024, GRB Coordinates Network, 37962, 1

\bibitem[{{Breeveld} {et~al.}(2024){Breeveld}, {Klingler}, \& {Swift/UVOT Team}}]{2024GCN.37974....1B}
{Breeveld}, A.~A., {Klingler}, N.~J., \& {Swift/UVOT Team}. 2024, GRB Coordinates Network, 37974, 1

\bibitem[{{Cash}(1979)}]{Cash1979ApJ}
{Cash}, W. 1979, \apj, 228, 939, \dodoi{10.1086/156922}

\bibitem[{Cohen {et~al.}(1997)Cohen, Katz, Piran, Sari, Preece, \& Band}]{cohenPossibleEvidenceRelativistic1997}
Cohen, E., Katz, J.~I., Piran, T., {et~al.} 1997, The Astrophysical Journal, 488, 330, \dodoi{10.1086/304699}

\bibitem[{{Daigne} {et~al.}(2011){Daigne}, {Bo{\v{s}}njak}, \& {Dubus}}]{2011A&A...526A.110D}
{Daigne}, F., {Bo{\v{s}}njak}, {\v{Z}}., \& {Dubus}, G. 2011, \aap, 526, A110, \dodoi{10.1051/0004-6361/201015457}

\bibitem[{{Derishev} {et~al.}(2001){Derishev}, {Kocharovsky}, \& {Kocharovsky}}]{2001A&A...372.1071D}
{Derishev}, E.~V., {Kocharovsky}, V.~V., \& {Kocharovsky}, V.~V. 2001, \aap, 372, 1071, \dodoi{10.1051/0004-6361:20010586}

\bibitem[{Feigelson \& Babu(2012)}]{Feigelson_Babu_2012}
Feigelson, E.~D., \& Babu, G.~J. 2012, Modern Statistical Methods for Astronomy: With R Applications (Cambridge University Press)

\bibitem[{{Fermi GBM Team}(2024)}]{2024GCN.37955....1F}
{Fermi GBM Team}. 2024, GRB Coordinates Network, 37955, 1

\bibitem[{{Gehrels} {et~al.}(2006){Gehrels}, {Norris}, {Barthelmy}, {Granot}, {Kaneko}, {Kouveliotou}, {Markwardt}, {M{\'e}sz{\'a}ros}, {Nakar}, {Nousek}, {O'Brien}, {Page}, {Palmer}, {Parsons}, {Roming}, {Sakamoto}, {Sarazin}, {Schady}, {Stamatikos}, \& {Woosley}}]{2006Natur.444.1044G}
{Gehrels}, N., {Norris}, J.~P., {Barthelmy}, S.~D., {et~al.} 2006, \nat, 444, 1044, \dodoi{10.1038/nature05376}

\bibitem[{{Ghirlanda} {et~al.}(2002){Ghirlanda}, {Celotti}, \& {Ghisellini}}]{2002A&A...393..409G}
{Ghirlanda}, G., {Celotti}, A., \& {Ghisellini}, G. 2002, \aap, 393, 409, \dodoi{10.1051/0004-6361:20021038}

\bibitem[{{Ghisellini} {et~al.}(2000){Ghisellini}, {Celotti}, \& {Lazzati}}]{2000MNRAS.313L...1G}
{Ghisellini}, G., {Celotti}, A., \& {Lazzati}, D. 2000, \mnras, 313, L1, \dodoi{10.1046/j.1365-8711.2000.03354.x}

\bibitem[{{Ghosh} {et~al.}(2024){Ghosh}, {Razzaque}, {Moskvitin}, \& {Sotnikova}}]{2024GCN.38220....1G}
{Ghosh}, A., {Razzaque}, S., {Moskvitin}, A., \& {Sotnikova}, Y. 2024, GRB Coordinates Network, 38220, 1

\bibitem[{{Gompertz} {et~al.}(2023){Gompertz}, {Ravasio}, {Nicholl}, {Levan}, {Metzger}, {Oates}, {Lamb}, {Fong}, {Malesani}, {Rastinejad}, {Tanvir}, {Evans}, {Jonker}, {Page}, \& {Pe'er}}]{2023NatAs...7...67G}
{Gompertz}, B.~P., {Ravasio}, M.~E., {Nicholl}, M., {et~al.} 2023, Nature Astronomy, 7, 67, \dodoi{10.1038/s41550-022-01819-4}

\bibitem[{Gruber {et~al.}(2014)Gruber, Goldstein, {Weller von Ahlefeld}, Narayana~Bhat, Bissaldi, Briggs, Byrne, Cleveland, Connaughton, Diehl, Fishman, Fitzpatrick, Foley, Gibby, Giles, Greiner, Guiriec, {van der Horst}, {von Kienlin}, Kouveliotou, Layden, Lin, Meegan, McGlynn, Paciesas, Pelassa, Preece, Rau, {Wilson-Hodge}, Xiong, Younes, \& Yu}]{gruberFermiGBMGammaRay2014}
Gruber, D., Goldstein, A., {Weller von Ahlefeld}, V., {et~al.} 2014, ApJS, 211

\bibitem[{Kaneko {et~al.}(2006)Kaneko, Preece, Briggs, Paciesas, Meegan, \& Band}]{kanekoCompleteSpectralCatalog2006}
Kaneko, Y., Preece, R.~D., Briggs, M.~S., {et~al.} 2006, ApJS, 166, 298, \dodoi{10.1086/505911}

\bibitem[{{Klingler} {et~al.}(2024){Klingler}, {Dichiara}, {Gupta}, {Palmer}, {Siegel}, \& {Neil Gehrels Swift Observatory Team}}]{2024GCN.37956....1K}
{Klingler}, N.~J., {Dichiara}, S., {Gupta}, R., {et~al.} 2024, GRB Coordinates Network, 37956, 1

\bibitem[{Kobayashi {et~al.}(1997)Kobayashi, Piran, \& Sari}]{kobayashiCanInternalShocks1997}
Kobayashi, S., Piran, T., \& Sari, R. 1997, The Astrophysical Journal, 490, 92, \dodoi{10.1086/512791}

\bibitem[{{Lipunov} {et~al.}(2024){Lipunov}, {Kornilov}, {Gorbovskoy}, {Zhirkov}, {Tyurina}, {Balanutsa}, {Kuznetsov}, {Senik}, {Vlasenko}, {Antipov}, {Zimnukhov}, {Minkina}, {Chasovnikov}, {Topolev}, {Kuvshinov}, {Cheryasov}, {Kechin}, {Tselik}, {Sosnovskij}, {Podesta}, {Lopez}, {Podesta}, {Francile}, {Rebolo}, {Serra}, {Buckley}, {Gress}, {Budnev}, {Ershova}, {Carrasco}, {Valdes}, {Chavushyan}, {Patino Alvarez}, {Martinez}, {Corella}, {Rodriguez}, {Tlatov}, {Dormidontov}, {Yurkov}, \& {Gabovich}}]{2024GCN.37975....1L}
{Lipunov}, V., {Kornilov}, V., {Gorbovskoy}, E., {et~al.} 2024, GRB Coordinates Network, 37975, 1

\bibitem[{Meegan {et~al.}(2009)Meegan, Lichti, Bhat, Bissaldi, Briggs, Connaughton, Diehl, Fishman, Greiner, Hoover, van~der Horst, von Kienlin, Kippen, Kouveliotou, McBreen, Paciesas, Preece, Steinle, Wallace, Wilson, \& Wilson-Hodge}]{Meegan2009}
Meegan, C., Lichti, G., Bhat, P.~N., {et~al.} 2009, The Astrophysical Journal, 702, 791, \dodoi{10.1088/0004-637X/702/1/791}

\bibitem[{{Nakar} {et~al.}(2009){Nakar}, {Ando}, \& {Sari}}]{2009ApJ...703..675N}
{Nakar}, E., {Ando}, S., \& {Sari}, R. 2009, \apj, 703, 675, \dodoi{10.1088/0004-637X/703/1/675}

\bibitem[{{Norris} {et~al.}(2000){Norris}, {Marani}, \& {Bonnell}}]{2000ApJ...534..248N}
{Norris}, J.~P., {Marani}, G.~F., \& {Bonnell}, J.~T. 2000, \apj, 534, 248, \dodoi{10.1086/308725}

\bibitem[{{Oganesyan} {et~al.}(2017){Oganesyan}, {Nava}, {Ghirlanda}, \& {Celotti}}]{2017ApJ...846..137O}
{Oganesyan}, G., {Nava}, L., {Ghirlanda}, G., \& {Celotti}, A. 2017, \apj, 846, 137, \dodoi{10.3847/1538-4357/aa831e}

\bibitem[{{Oganesyan} {et~al.}(2018){Oganesyan}, {Nava}, {Ghirlanda}, \& {Celotti}}]{2018A&A...616A.138O}
---. 2018, \aap, 616, A138, \dodoi{10.1051/0004-6361/201732172}

\bibitem[{Panaitescu \& Kumar(2002)}]{panaitescuPropertiesRelativisticJets2002}
Panaitescu, A., \& Kumar, P. 2002, The Astrophysical Journal, 571, 779

\bibitem[{{Pe'er} \& {Zhang}(2006)}]{2006ApJ...653..454P}
{Pe'er}, A., \& {Zhang}, B. 2006, \apj, 653, 454, \dodoi{10.1086/508681}

\bibitem[{Piran(1999)}]{piranGammarayBurstsFireball1999}
Piran, T. 1999, Physics Reports, 314, 575

\bibitem[{{Preece} {et~al.}(1998){Preece}, {Briggs}, {Mallozzi}, {Pendleton}, {Paciesas}, \& {Band}}]{1998ApJ...506L..23P}
{Preece}, R.~D., {Briggs}, M.~S., {Mallozzi}, R.~S., {et~al.} 1998, \apjl, 506, L23, \dodoi{10.1086/311644}

\bibitem[{Ravasio {et~al.}(2019)Ravasio, Ghirlanda, Nava, \& Ghisellini}]{ravasioEvidenceTwoSpectral2019}
Ravasio, M.~E., Ghirlanda, G., Nava, L., \& Ghisellini, G. 2019, Astronomy and Astrophysics, 625

\bibitem[{{Ravasio} {et~al.}(2018){Ravasio}, {Oganesyan}, {Ghirlanda}, {Nava}, {Ghisellini}, {Pescalli}, \& {Celotti}}]{2018A&A...613A..16R}
{Ravasio}, M.~E., {Oganesyan}, G., {Ghirlanda}, G., {et~al.} 2018, \aap, 613, A16, \dodoi{10.1051/0004-6361/201732245}

\bibitem[{{Sari}(1998)}]{1998ApJ...494L..49S}
{Sari}, R. 1998, \apjl, 494, L49, \dodoi{10.1086/311160}

\bibitem[{Sari \& Piran(1995)}]{sariHydrodynamicTimescalesTemporal1995}
Sari, R., \& Piran, T. 1995, The Astrophysical Journal, 455, L143, \dodoi{10.1086/309835}

\bibitem[{{Sari} {et~al.}(1998){Sari}, {Piran}, \& {Narayan}}]{1998ApJ...497L..17S}
{Sari}, R., {Piran}, T., \& {Narayan}, R. 1998, \apjl, 497, L17, \dodoi{10.1086/311269}

\bibitem[{Scargle {et~al.}(2013)Scargle, Norris, Jackson, \& Chiang}]{Scargle2013}
Scargle, J.~D., Norris, J.~P., Jackson, B., \& Chiang, J. 2013, The Astrophysical Journal, 764, 167, \dodoi{10.1088/0004-637X/764/2/167}

\bibitem[{Schwarz(1978)}]{bic_ref}
Schwarz, G. 1978, The Annals of Statistics, 6, 461 , \dodoi{10.1214/aos/1176344136}

\bibitem[{Shen {et~al.}(2005)Shen, Song, \& Li}]{shenSpectralLagsEnergy2005}
Shen, R.-F., Song, L.-M., \& Li, Z. 2005, Monthly Notices of the Royal Astronomical Society, 362, 59, \dodoi{10.1111/j.1365-2966.2005.09163.x}

\bibitem[{Sironi \& Spitkovsky(2011)}]{sironiPARTICLEACCELERATIONRELATIVISTIC2011}
Sironi, L., \& Spitkovsky, A. 2011, The Astrophysical Journal, 726, 75, \dodoi{10.1088/0004-637X/726/2/75}

\bibitem[{Sironi {et~al.}(2013)Sironi, Spitkovsky, \& Arons}]{sironiMaximumEnergyAccelerated2013}
Sironi, L., Spitkovsky, A., \& Arons, J. 2013, The Astrophysical Journal, 771

\bibitem[{Uhm \& Zhang(2014)}]{uhmFastcoolingSynchrotronRadiation2014}
Uhm, Z.~L., \& Zhang, B. 2014, Nature Physics, 10, 351, \dodoi{10.1038/nphys2932}

\bibitem[{{Uhm} \& {Zhang}(2014)}]{2014NatPh..10..351U}
{Uhm}, Z.~L., \& {Zhang}, B. 2014, Nature Physics, 10, 351, \dodoi{10.1038/nphys2932}

\bibitem[{{Uhm} \& {Zhang}(2016)}]{2016ApJ...825...97U}
---. 2016, \apj, 825, 97, \dodoi{10.3847/0004-637X/825/2/97}

\bibitem[{Uhm \& Zhang(2016)}]{uhmUnderstandingGRBPrompt2016}
Uhm, Z.~L., \& Zhang, B. 2016, The Astrophysical Journal, 825

\bibitem[{Ukwatta {et~al.}(2010)Ukwatta, Stamatikos, Dhuga, Sakamoto, Barthelmy, Eskandarian, Gehrels, Maximon, Norris, \& Parke}]{Ukwatta_2010}
Ukwatta, T.~N., Stamatikos, M., Dhuga, K.~S., {et~al.} 2010, The Astrophysical Journal, 711, 1073, \dodoi{10.1088/0004-637X/711/2/1073}

\bibitem[{{Wang} {et~al.}(2025){Wang}, {Zhou}, {Wang}, {Ren}, {Tinyanont}, {Xu}, {Sun}, {Fynbo}, {Malesani}, {An}, {Anutarawiramku}, {Butpa}, {Fu}, {Jiang}, {Liu}, {Palee}, {Prasit}, {Zhu}, {Jin}, \& {Wei}}]{2025arXiv250104906W}
{Wang}, Q.-L., {Zhou}, H., {Wang}, Y., {et~al.} 2025, arXiv e-prints, arXiv:2501.04906, \dodoi{10.48550/arXiv.2501.04906}

\bibitem[{Wu \& Fenimore(2000)}]{wuSpectralLagsGammaRay2000}
Wu, B., \& Fenimore, E. 2000, The Astrophysical Journal, 535, L29, \dodoi{10.1086/312700}

\bibitem[{Yost {et~al.}(2003)Yost, Harrison, Sari, \& Frail}]{yostStudyAfterglowsFour2003}
Yost, S.~A., Harrison, F.~A., Sari, R., \& Frail, D.~A. 2003, The Astrophysical Journal, 597, 459, \dodoi{10.1086/378288}

\bibitem[{Zhang \& Yan(2011)}]{ZhangYan2011}
Zhang, B., \& Yan, H. 2011, The Astrophysical Journal, 726, \dodoi{10.1088/0004-637X/726/2/90}

\bibitem[{{Zhang} \& {Yan}(2011)}]{2011ApJ...726...90Z}
{Zhang}, B., \& {Yan}, H. 2011, \apj, 726, 90, \dodoi{10.1088/0004-637X/726/2/90}

\bibitem[{{Zhang} {et~al.}(2009){Zhang}, {Zhang}, {Virgili}, {Liang}, {Kann}, {Wu}, {Proga}, {Lv}, {Toma}, {M{\'e}sz{\'a}ros}, {Burrows}, {Roming}, \& {Gehrels}}]{2009ApJ...703.1696Z}
{Zhang}, B., {Zhang}, B.-B., {Virgili}, F.~J., {et~al.} 2009, \apj, 703, 1696, \dodoi{10.1088/0004-637X/703/2/1696}

\bibitem[{Zhang {et~al.}(2012)Zhang, Burrows, Zhang, Mészáros, Wang, Stratta, D'Elia, Frederiks, Golenetskii, Cummings, Norris, Falcone, Barthelmy, \& Gehrels}]{Zhang_2012}
Zhang, B.-B., Burrows, D.~N., Zhang, B., {et~al.} 2012, The Astrophysical Journal, 748, 132, \dodoi{10.1088/0004-637X/748/2/132}

\bibitem[{{Zhao} {et~al.}(2014){Zhao}, {Li}, {Liu}, {Zhang}, {Bai}, \& {M{\'e}sz{\'a}ros}}]{2014ApJ...780...12Z}
{Zhao}, X., {Li}, Z., {Liu}, X., {et~al.} 2014, \apj, 780, 12, \dodoi{10.1088/0004-637X/780/1/12}

\bibitem[{Zheng {et~al.}(2024)}]{GCN37959}
Zheng, W., {et~al.} 2024, {GRB 241030A: Keck/LRIS spectroscopic redshift z = 1.411}, \url{https://gcn.nasa.gov/circulars/37959}

\end{thebibliography}

\end{document}